\documentclass[reprint]{JASA}

\usepackage{tikz,tikz-qtree}	
\usepackage{graphicx}
\usepackage{amsmath,amssymb,bbm,enumitem}
\usetikzlibrary{calc,patterns,angles,quotes}
\usepackage{mathrsfs}
\usepackage{mathtools}
\usepackage{setspace} 
\usepackage[normalem]{ulem}
\usepackage{rotating}
\tikzset{ 
 lspk/.pic={\draw (0,0) rectangle (.3,.1);
 \draw (-.1,-0.1) -- (.4,-0.1);
 \draw (0,0) -- (-.1,-.1);
 \draw (0.3,0) -- (0.4,-.1);
 }} 
\tikzset{
    man/.pic={\draw (0,0) circle (.2);
    \draw (0,0.3) -- (-.1,0.2);
    \draw (0,0.3) -- (0.1,0.2);}
     }

\newcommand{\localized}[1]{
\begin{tikzpicture}[scale=.5]
 \draw (2-2.82,1+2.82) pic[rotate=20] {lspk}; 
 \draw (2,1) pic {man};

\draw[dashed] (2-2.82+.1,1+2.82-.1) -- (2-.2,1 ); 
\draw[dashed] (2-2.82+.1,1+2.82-.1) -- (2+.2,1 ); 
\draw[dashed] (2-2.82+.1,1+2.82-.1) -- (2,1) ; 

\draw[dashed] (2-2.82,1+2.82) -- (2,1+2.82 ); 
\draw[dashed] (2,1) -- (2,1+2.82 ); 

\node[text width=1cm, anchor=west, left] at (2.5,2){$\theta_m$};
\draw (2,2) arc (90:132:1);

\node[text width=1cm, anchor=west, left] at (2,1){\footnotesize{$x_l[n]$}};
\node[text width=1cm, anchor=west, right] at (2.25,1){\footnotesize{$x_r[n]$}};

\end{tikzpicture}}

\newcommand{\wider}[1]{ 
\begin{tikzpicture}[scale=.5]
\draw (2-1.03,1+3.86) pic[rotate=30] {lspk}; 
\draw (2-2,1+3.46) pic[rotate=25] {lspk}; 
\draw (2-2.82,1+2.82) pic[rotate=20] {lspk}; 
\draw (2-3.46,1+2) pic[rotate=25] {lspk}; 
\draw (2-3.86,1+1.03) pic[rotate=30] {lspk}; 
\draw (2,1) pic {man};

\draw[dashed] (2-1.03+.1,1+3.86-.1) -- (2-.2,1 ); 
\draw[dashed] (2-1.03+.1,1+3.86-.1) -- (2+.2,1 ); 
\draw[dashed] (2-3.86+.1,1+1.03-.1) -- (2-.2,1 ); 
\draw[dashed] (2-3.86+.1,1+1.03-.1) -- (2+.2,1 ); 

\node[text width=1cm, anchor=west, left] at (1.5,2){$\alpha$};
\draw (2-.65,1+2.42) arc (90+15:90+75:2.5); 

\draw[dashed] (2-2.82+.1,1+2.82-.1) -- (2,1) ; 

\draw[dashed] (2,1) -- (2,1+2.82 ); 

\node[text width=1cm, anchor=west, left] at (2.5,2){$\theta_m$};
\draw (2,2) arc (90:132:1);

\end{tikzpicture}}

\newcommand{\reverb}[1]{ 
\begin{tikzpicture}[scale=.5]
\draw (2-1.03,1+3.86) pic[rotate=30] {lspk}; 
\draw (2-2,1+3.46) pic[rotate=25] {lspk}; 
\draw (2-2.82,1+2.82) pic[rotate=20] {lspk}; 
\draw (2-3.46,1+2) pic[rotate=25] {lspk}; 
\draw (2-3.86,1+1.03) pic[rotate=30] {lspk}; 
\draw (2,1) pic {man};

\draw[dashed] (2-1.03+.1,1+3.86-.1) -- (2-.2,1 ); 
\draw[dashed] (2-1.03+.1,1+3.86-.1) -- (2+.2,1 ); 
\draw[dashed] (2-3.86+.1,1+1.03-.1) -- (2-.2,1 ); 
\draw[dashed] (2-3.86+.1,1+1.03-.1) -- (2+.2,1 ); 

\node[text width=1cm, anchor=west, left] at (1.5,2){$\alpha$};
\draw (2-.65,1+2.42) arc (90+15:90+75:2.5); 

\draw[dashed] (2-2.82+.1,1+2.82-.1) -- (2,1) ; 

\draw[dashed] (2,1) -- (2,1+2.82 ); 

\node[text width=1cm, anchor=west, left] at (2.5,2){$\theta_m$};
\draw (2,2) arc (90:132:1);

\end{tikzpicture}}
\newcommand{\reverbmore}[1]{ 
\begin{tikzpicture}[scale=.5]
\draw (2-1.03,1+3.86) pic[rotate=30] {lspk}; 
\draw (2-2,1+3.46) pic[rotate=25] {lspk}; 
\draw (2-2.82,1+2.82) pic[rotate=20] {lspk}; 
\draw (2-3.46,1+2) pic[rotate=25] {lspk}; 
\draw (2-3.86,1+1.03) pic[rotate=30] {lspk}; 
\draw (2,1) pic {man};

\draw[dashed] (2-1.03+.1,1+3.86-.1) -- (2-.2,1 ); 
\draw[dashed] (2-1.03+.1,1+3.86-.1) -- (2+.2,1 ); 
\draw[dashed] (2-3.86+.1,1+1.03-.1) -- (2-.2,1 ); 
\draw[dashed] (2-3.86+.1,1+1.03-.1) -- (2+.2,1 ); 

\node[text width=1cm, anchor=west, left] at (1.5,2){$\alpha$};
\draw (2-.65,1+2.42) arc (90+15:90+75:2.5); 

\draw[dashed] (2-2.82+.1,1+2.82-.1) -- (2,1) ; 

\draw[dashed] (2,1) -- (2,1+2.82 ); 

\node[text width=1cm, anchor=west, left] at (2.5,2){$\theta_m$};
\draw (2,2) arc (90:132:1);

\end{tikzpicture}}
\newcommand{\surround}{\begin{tikzpicture}[scale=.25]

\draw (2-0,1+4) pic {lspk}; 
\draw (2-2.82,1+2.82) pic[rotate=25] {lspk}; 
\draw (2-2.82,1-2.82) pic[rotate=90+25] {lspk}; 
\draw (2-0,1-4) pic[rotate=180] {lspk}; 
\draw (2+4,1+0) pic[rotate=-90] {lspk}; 
 \draw (2,1) pic {man};

\draw[dashed] (2-2.82+.1,1+2.82-.1) -- (2,1) ; 

\end{tikzpicture}}
\newcommand{\sigloc}{\begin{tikzpicture}[scale=.25]

\draw[->] (-4,0) -- (4,0) node[right]{$\theta$} ; 
\draw[->] (0,0) -- (0,5) node[left]{$\sigma^2$} ; 

\draw (3,0) -- (3,4) ;  

\node[text width=1cm, anchor=west, right] at (0-0.5,0-1){0};
\node[text width=1cm, anchor=west, right] at (2.5,0-1){$\theta_m$};
\end{tikzpicture}}

\newcommand{\sigwide}{\begin{tikzpicture}[scale=.25]

\draw[->] (-3,0) -- (6,0) node[right]{$\theta$} ; 
\draw[->] (0,0) -- (0,5) node[left]{$\sigma^2$} ; 

\draw (1,0) -- (1,4) ; 
\draw (2,0) -- (2,4) ; 
\draw (3,0) -- (3,4) ;  
\draw (4,0) -- (4,4) ; 
\draw (5,0) -- (5,4) ; 

\node[text width=1cm, anchor=west, right] at (0-0.5,0-1){0};
\node[text width=1cm, anchor=west, right] at (2.5,0-1){$\theta_m$};
\end{tikzpicture}}

\newcommand{\sigreverb}{\begin{tikzpicture}[scale=.25]

\draw[->] (-4,0) -- (6,0) node[right]{$\theta$}; 
\draw[->] (0,0) -- (0,5) node[left]{$\sigma^2$}; 

\draw (1,0) -- (1,2) ; 
\draw (2,0) -- (2,2) ; 
\draw (3,0) -- (3,4) ;  
\draw (4,0) -- (4,2) ; 
\draw (5,0) -- (5,2) ; 

\node[text width=1cm, anchor=west, right] at (0-0.5,0-1){0};
\node[text width=1cm, anchor=west, right] at (3,0-1){$\theta_m$};
\end{tikzpicture}}

\newcommand{\sigsurround}{\begin{tikzpicture}[scale=.25]

\draw[->] (-6,0) -- (6,0) node[right]{$\theta$}; 
\draw[->] (0,0) -- (0,5) node[left]{$\sigma^2$} ; 

\draw (-5,0) -- (-5,1) ; 
\draw (-4,0) -- (-4,1) ; 
\draw (3,0) -- (3,4) ;  
\draw (5,0) -- (5,1) ; 
\draw[->] (0,0) -- (0,1) ; 

\node[text width=1cm, anchor=west, right] at (0-0.5,0-1){0};
\node[text width=1cm, anchor=west, right] at (2.5,0-1){$\theta_m$};
\end{tikzpicture}}

\tikzset{ 
 lspk/.pic={\draw (0,0) rectangle (.3,.1);
 \draw (-.1,-0.1) -- (.4,-0.1);
 \draw (0,0) -- (-.1,-.1);
 \draw (0.3,0) -- (0.4,-.1);
 }} 
\tikzset{
    man/.pic={\draw (0,0) circle (.2);
    \draw (0,0.3) -- (-.1,0.2);
    \draw (0,0.3) -- (0.1,0.2);}
     }
\newcommand{\vswA}[1]{
\begin{tikzpicture}[scale=0.5]

\draw (-.7,.7) circle (.1);
\draw (-0.5,0.86) circle (.1);
\draw (-.86,0.5) circle (.1);
\draw (-0.26,0.96) circle (.1);
\draw (-0.96,0.26) circle (.1);
\draw (0,0) pic {man};
\draw[dotted] (0,0) circle (1.5); 
\end{tikzpicture}}

\newcommand{\vswB}[1]{
\begin{tikzpicture}[scale=0.5]
\draw (0,0) pic {man};

\draw (-1,0) circle (.1);
\draw (-0.86,0.5) circle (.1);
\draw (-0.5,0.86) circle (.1);
\draw (0,1) circle (.1);
\draw (0.5,0.86) circle (.1);
\draw (0.86,0.5) circle (.1);
\draw (1,0) circle (.1);
\draw (0.86,-0.5) circle (.1);
\draw (0.5,-0.86) circle (.1);
\draw (0,-1) circle (.1);
\draw (-0.5,-0.86) circle (.1);
\draw (-0.86,-0.5) circle (.1);

\draw[dotted] (0,0) circle (1.5); 

\end{tikzpicture}}

\newcommand{\reverbA}[1]{
\begin{tikzpicture}[scale=0.5]
\draw (0,0) pic {man};
\draw (0,0) circle (1.5);

\draw (-0.7,0.7) circle (.1);

\draw[dotted,->]  (-0.7,0.7) --  (0.75,1.3);
\draw[dotted,->]  (0.75,1.3) -- (-.2,0) ;
\draw[dotted,->]  (0.75,1.3) -- (+.2,0) ;

\draw[->]  (-0.7,0.7) --  ( 0.2,0 );
\draw[->]  (-0.7,0.7) -- (-.2,0) ;

\draw (-0.7,0.7) circle (.1);

\end{tikzpicture}}

\newcommand{\reverbB}[1]{
\begin{tikzpicture}[scale=0.5]
\draw (0,0) pic {man};
\draw (0,0) circle (1.5); 

\draw (-0.7,0.7) circle (.1);

\draw[->]  (-0.7,0.7) --  ( 0.2,0 );
\draw[->]  (-0.7,0.7) -- (-.2,0) ;

\draw[dotted,->]  (-0.7,0.7) --  (0.75,1.3);
\draw[dotted,->]  (0.75,1.3) -- (-.2,0) ;
\draw[dotted,->]  (0.75,1.3) -- (+.2,0) ;

\draw[dotted,->]  (-0.7,0.7) --  (-0.75,-1.3);
\draw[dotted,->]  (-0.75,-1.3) -- (-.2,0) ;
\draw[dotted,->]  (-0.75,-1.3) -- (+.2,0) ;

\draw[dotted,->]  (-0.7,0.7) --  (1.3,-0.75);
\draw[dotted,->]  (1.3,-0.75) -- (-.2,0) ;
\draw[dotted,->]  (1.3,-0.75) -- (+.2,0) ;

\end{tikzpicture}}

\tikzset{ 
 lspk/.pic={\draw (0,0) rectangle (.3,.1);
 \draw (-.1,-0.1) -- (.4,-0.1);
 \draw (0,0) -- (-.1,-.1);
 \draw (0.3,0) -- (0.4,-.1);
 }} 
\tikzset{
    man/.pic={\draw (0,0) circle (.2);
    \draw (0,0.3) -- (-.1,0.2);
    \draw (0,0.3) -- (0.1,0.2);}
     }
\newcommand{\reverbvsw}[1]{
\begin{tikzpicture}[scale=1]
\node[text width=1cm, left] at (3,-0.5){RSW};
\node[text width=1cm, right] at (4,-0.5){LEV};
\node[text width=1cm, below] at (0,0){Point source};
\node[text width=1cm, anchor=west, left] at (2,2){ESW};
\node[text width=2cm, anchor=west, left] at (4,3){ESW-$360^0$};
\draw[->]  (0,0) --  (4,0);
\node at (0,0) (A){};
\node at (4,3) (B){};
\node at (-3,3.5) (C){};
\node[right, above,align=center] at (3.5,0){RSW \\ Morimoto continuum
 };
\draw[dashed,->]  (A) --  node[midway, below, sloped]{proposed ESW continuum}(B);
\draw[dashed,->]  (A) --  node[midway, below, sloped]{proposed DSW continuum}(C);
\draw[dashed,->]  (A) -- node[midway, above, sloped]{Ensemble}(B);
\draw[dashed,->]  (A) -- node[midway, above, sloped]{Hearing impairment}(C);
\end{tikzpicture}
}

\allowdisplaybreaks[1]
\begin{document}

\title[JASA/Measure for source width]{Spatiogram: A phase based directional angular measure and perceptual weighting for ensemble source width}
\author{Arthi S}
\author{T V Sreenivas}
\affiliation{Indian Institute of Science, Bangalore-12, India}



\begin{abstract}
In concert hall studies, inter-aural cross-correlation (IACC), which is signal dependent, is used as a measure of perceptual  source width. The same measure is used for perceptual source width in the case of distributed sources also.  In this work, we examine the validity of IACC for both the cases and develop an improved measure for ensemble-like distributed sources. We decompose the new objective measure for perceptual ensemble source width (ESW) into two components (i) phase based directional angular measure, which is timbre independent (spatial measure) and (ii) mean time-bandwidth energy (MTBE), a perceptual weight, (timbre measure). This combination of spatial and timbral measures can  be extended as an alternate measure for determining auditory source width (ASW) and listener envelopment (LEV) of arbitrary signals in concert-hall and room acoustics.
\end{abstract}


\maketitle

\section{\label{sec:1} Introduction}
In audio rendering, recreating the acoustic event with naturalness of the source timbre and the  environmental characteristics is required along with the fidelity of signal rendering. In concert hall acoustics, natural reverberation is created through the acoustic design for effective perceptual experience of listeners. However, in virtual reality applications, it is important to create through signal processing, various spatial attributes of a given sound such as its direction, size, envelopment, etc. In this work, focusing on the source signal attributes, we study the perceptual size/width of a source as perceived by a listener. A wide road with vehicles moving, waterfalls, partially or fully filled large gallery, an orchestra of many instruments, rumbling of tree leaves in a forest, etc are some of the examples of sources in natural scenario with a perceived source width! In contrast, a localised source would be a bird call or human speech. In the scheme of sound rendering, a sound object can be presented through several distributed sources of similar spectral distribution to achieve an integrated perception or cognition which gives rise to a single whole wide auditory object perception. We will refer to this type of rendering, resulting in  a virtual wide source as ``ensemble source width'' (ESW). 
\par 
In the literature, (1-IACC) is used as a measure of perceived source width extent \cite{nowak2013perception}  \cite{hirvonen2006perception}, which has been derived mainly from the concert-hall studies. In the present work, we examine the validity of this measure for ensemble-like presentation and propose a new measure which is a combination of (i) phase based directional angular measure, which is independent of signal timbre and (ii) a perceptual scalar weight, called mean time-bandwidth energy (MTBE), which is dependent on the signal timbre; This combination is observed as a better measure of source width perception in this work.
\par 
\vspace*{-2mm}
\subsection{Measure of ASW and LEV} 
Traditionally, IACC has been observed to correlate with the source widening in the design of concert hall acoustics\cite{morimoto1995practical}. Given $x_l(t)$ and $x_r(t)$ as the left ear and right ear signals, IACC is defined ideally as \cite{blauert1997spatial}:
\begin{equation}
 \phi(\tau) =  \lim_{T->\infty}\frac{\frac{1}{2T}\int_{-T}^{+T}x_l(t)x_r(t+\tau)dt}{\frac{1}{2T}\sqrt{\int_{-T}^{T}x_l^2(t)dt}\sqrt{\int_{-T}^{T}x_r^2(t)dt}}
\end{equation}
\begin{equation}
  IACC \:\: \Phi =  \max_{\tau} \phi(\tau)
\end{equation}
Here,  $x_l(t)$ and $x_r(t)$ are defined through linear models of acoustic signal propagation using room impulse response (RIR) and head related impulse response (HRIR).
\begin{align*}
x_l(t) = s(t) * r(t)* h_l(t) \\
x_t(t) = s(t) * r(t)* h_r(t) 
\end{align*}
We observe that IACC is dependent on signal s(t), location dependent room impulse response r(t) and  also listener dependent HRIR of the measurement device $h_r(t) and h_l(t)$ \cite{morimoto1995practical}. Presently, standardizing the source signal ( Mozart Symphony No. 41 ``Jupiter'') and the measurement device, the concert halls are characterized using IACC \cite{morimoto1995practical}\cite{barron1981spatial}. Lateral fraction (LF) energy is another promising measure for auditory source width (ASW) of wide band signals \cite{barron1981spatial}\cite{morimoto2005appropriate}. It is defined as the ratio of  lateral direction energy to omni directional energy and we observe that again it is a signal and location dependent measure. \par
While ASW is one aspect of spatial impression, listener envelopment (LEV) is another aspect, where listener feels immersed in the acoustic environment. LEV is dependent on many factors like signal strength, early and late reverberation, etc and  angular distribution of late reverberation is found to play a significant role \cite{bradley1995objective}\cite{morimoto2001role}. A combination of lateral fraction energy and IACC  has been proposed as an objective measure of LEV \cite{soulodre2004new}. 
\par
The above studies are mainly from concert hall acoustics and in the recent studies of perception of distributed sources, IACC has been studied as an objective measure to predict perceptual width. In the experimental studies of multi-channel rendering \cite{hirvonen2006perception},  direction perception of distributed sources of different sub-band frequencies (200-1800 Hz) is studied. In correlating the perception and signal parameters using binaural correlogram, the signal magnitude and HRIR magnitude of multiple sources create multiple peaks and interpretation of maximum of IACC function is difficult in multi-source case \cite{hirvonen2006perception}. These correlogram based measures show limitations of IACC and sub-band IACC as a measure of physical extent, because of interfering magnitude effects in modeling distributed source perception. While the measure of de-correlation is an intuitive measure for relative reverberation, standardizing the source signal is suggested. Thus, IACC is still signal dependent and not effective as a  signal independent measure even for the case of reverberation  \cite{morimoto2005appropriate}. 
\par
We observe from the above discussion that in all these measure of ASW and LEV in concert hall acoustics and distributed sources, the objective measures are signal dependent and measure to predict source width for arbitrary signals is still an open question and also desirable from an engineering point of view, particularly in the context of virtual reality (VR) and ensemble rendering applications. 
\par
The above measures of IACC and LF  are stationary measures. There has been an extensive effort to obtain an objective measure to correlate perceptual source width to signal properties in which time-varying properties have  been explored. The spectro-temporal distribution of IACC(t,f) has  been studied and it is suggested that the derived metrics from IACC(t,f) could be of importance for reliable rendering of a source\cite{jackson2008qestral}. Inter-aural time difference (ITD) fluctuation as an objective measure of spatial impression has also been studied \cite{griesinger1999objective}\cite{mason2001interaural}. These measures of variation in IACC and ITD, similar to previous measure of IACC and LF, are also signal dependent, thus limiting their ability to compare the source widening across different signal timbres and different measurement devices (KEMAR). 
\par 
We also note that there is significant difference in the perception of  the width of a signal with reverberation and an ensemble source width (ESW). While reverberation is because of ambiance characteristics, ensemble width is a source attribute. In the case of reverberation, it is observed that while the direction of dominant component is salient perceptually, the direction of reflection components ($\leq -6$dB with respect to dominant component) are not significant \cite{morimoto1993relation}. Perceptually, we are sensitive mainly to the binaural cross-correlations, not to the directions of attenuated  components. But, this may not be the case in ensemble width perception. In the rendering of a distributed source, all the source signals would have nearly the same signal strength; hence, we could be sensitive to the direction of almost all the source signals, specifically the spatial distribution of the edges or  the spatial extrema of the sources \cite{santala2011directional}. 
\vspace*{-3mm}
\subsection{Dimensions of Auditory Source Width}
Perceptual source widening is observed in three scenarios (i) reverberation source width (RSW) (ii) Ensemble source width (ESW) (iii) Diffused source width (DSW) that is observed in hearing impaired listeners. In all these three cases, the physics behind the binaural signal modeling could be different and hence the  nature of signal timbre and nature/ degree of perceptual source width also differs. Hence, we consider three different perceptual dimensions corresponding to each physical process as illustrated in Fig.\ref{fig_reverbVSW}. The nomenclature, proposed by us, is summarized in Table.\ref{tab_nomenclature}. Morimoto et. al proposed one dimensional perceptual scale of spatial impression where a localized  source is a point source, followed by sources   having a  perceptual width due to reverberation (RSW). With increased angular distribution of reflections, it gives rise to envelopment (LEV) \cite{morimoto2001role} as shown in Fig.\ref{fig_reverbVSW} and Fig.\ref{fig_continuum}. 

\begin{table} [!b]
\vspace{-2mm}
\centerline{
\footnotesize{
\begin{tabular}{|p {2 cm}|p {2 cm}|p {3 cm}|}
\hline 
Nomenclature & Enclosure & Sound Source  \\
\hline
\hline
Apparent Source width 
or 
Auditory Source width (ASW)  & Common term for perceptual source widening & Single or distributed sources \\ 
\hline
Reverb source width (RSW) &  Reverb+ HRTF & Single source with multiple reflections  \\ 
\hline
Listener Envelopment (LEV) &  Reverb+ HRTF & Single source  with multiple reflections with increased angular distribution \\ 
\hline
Ensemble source width (ESW) & Anechoic+ HRTF & Distributed sources but integrated single source perception \\
\hline
Diffused source width (DSW) & Hearing impairment in presence of interference sources &  Asynchronous neuronal firing or reduced sensitivity to IACC could be a reason \\
\hline
\end{tabular}}}
\caption{{Nomenclature in spatial rendering of sound}}
\label{tab_nomenclature}
\end{table}
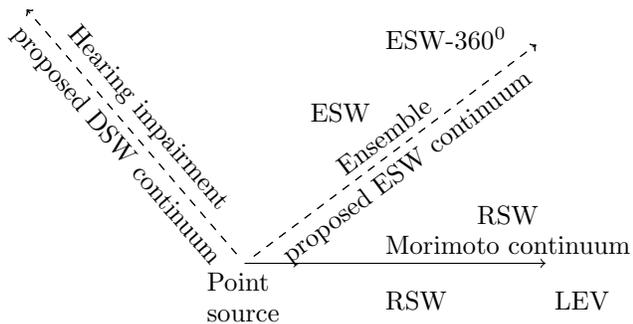
\begin{figure} [h!]
\begin{minipage}{1\linewidth}
\begin{tikzpicture}[scale=0.5]
    \node at (0,0) (test2) {\reverbvsw{}};
\end{tikzpicture}  
\end{minipage}
\caption{{Proposed ASW continuums (a) Reverb Source Width (RSW)  or Morimoto continuum \cite{morimoto2001role} (b) Ensemble source width (ESW) continuum (c) Diffused source width (DSW) continuum }}
\label{fig_reverbVSW}
\end{figure}
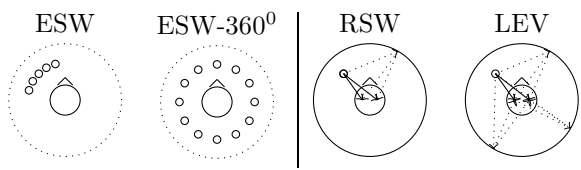
\begin{figure}  [h!]
\begin{minipage}{.22\linewidth}
\centering{ {ESW}} \\  
\begin{tikzpicture}[scale=1]
    \node at (0,0) (test2) {\vswA{}};
\end{tikzpicture}  
\end{minipage}
\begin{minipage}{.22\linewidth}
\centering{ {ESW-$360^0$}} \\ 
\begin{tikzpicture}[scale=1]
    \node at (0,0) (test2) {\vswB{}};
\end{tikzpicture}  
\end{minipage} 
\vrule
\begin{minipage}{.22\linewidth}
\centering{ {RSW}} \\ 
\begin{tikzpicture}[scale=1]
    \node at (0,0) (test2) {\reverbA{}};
\end{tikzpicture}  
\end{minipage} 
\begin{minipage}{.22\linewidth}
\centering{ {LEV}}  \\ 
\begin{tikzpicture}[scale=1]
    \node at (0,0) (test2) {\reverbB{}};
\end{tikzpicture}  
\end{minipage} \\ 
\caption{Illustration of proposed ESW continuum and Morimoto continuum \cite{morimoto2001role}}
\label{fig_continuum}
\end{figure}
\par
While a correlation based scalar measure such as IACC for ASW in the case of reverberation \cite{morimoto2005appropriate} is still an open question, in the present work, we develop a phase based angular measure, using correlation functions of HRIR, with motivation to localize as many sources. With known HRIR of a measurement device (KEMAR), we estimate spatial correlation between binaural GCC-PHAT and HRTF GCC-PHAT functions. We observe peaks corresponding to each directional source present in the ensemble, in this spatial correlation and the angular difference between the extreme sources is defined as the ``Angular width`` of the source. A similar approach could help in developing an objective measure for sources with reverberation by capturing the spatial distribution of early and late reverberation components, provided we can resolve binaurally the attenuated early and late components  in the estimation technique. 
\vspace*{-2mm}
\subsection{Timbre Vs Spatialisation}
Perceptual sensitivity of signals of different timbre for spatialisation has been studied using sub-band signals \cite{santala2011directional} \cite{otani2017largeness} \cite{mason2005frequency} and natural signals \cite{pihlajamaki2014synthesis}. Lower and higher frequencies have higher sensitivity to source width than mid-range of frequencies\cite{mason2005frequency}. Sensitivity to bandwidth and center frequency in the perception  of sound image has also been reported \cite{otani2017largeness} \cite{santala2011directional}. In these experiments, it is observed that there is high sensitivity of source width to low frequencies and wide bandwidths. In \cite{pihlajamaki2014synthesis}, different de-correlation methods and rendering techniques are studied for source width perception using natural signals of different timbre. HH
The authors observe that sustained signal instruments such as trombone, cello and guitar give rise to a wider source extent and transient sounds such as drums show poorer quality reproduction. These observations indicate that some signals are naturally better suited for a wide orchestral rendering, and some other instrument signals may not be so.
\par
In the present experimental work, we simulate an ensemble of different classes of instruments of same physical angular extent: wind, string, vocal  and percussion instruments of different spectro-temporal properties. 
More details of pitch and timbre variations across instruments is reported in \cite{arthi2019multi}.
\vspace*{-3mm}
\subsection{Present contribution}
We propose an objective measure for ESW using two factors (i) A phase based angular measure and (ii) a timbre-dependent perceptual weighting. We develop a spatial measure using spatial transform for multi-source localisation with personalised binaural HRIR correlation functions as functional bases. We experimentally verify relative perceptual measure of ensemble of different instruments correlate with timbre dependent measure. We develop a MUSHRA-like experimental methodology to determine the relative ensemble source width using  blind listening tests. This decomposition of spatial measure and timbral measure could be possibly extended to arbitrary signals with reverberation also.  
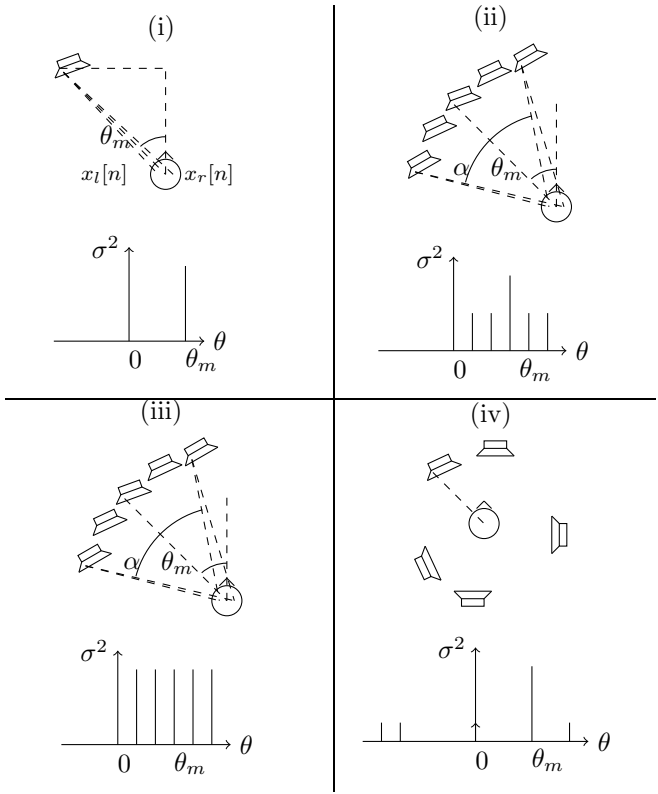
\begin{figure} [!h]
\centering
\begin{minipage}{0.48\linewidth}
\centering{(i)} \qquad \\
\begin{tikzpicture}[scale=0.6]
   \node at (0,0) (test1) {\localized{}};
    \node at (0,-4) (test1) {\sigloc{}};
\end{tikzpicture} 
\end{minipage}
\hfill
\vrule
\begin{minipage}{0.48\linewidth}
\centering{(ii)} \qquad \\
\begin{tikzpicture}[scale=0.6]
   \node at (0,0) (test3) {\reverb{}};
   \node at (0,-4) (test2) {\sigreverb{}};
\end{tikzpicture} \qquad \\
\end{minipage}
\hfill
\hrule
\vfill
\begin{minipage}{0.48\linewidth}
\centering{(iii)} \qquad \\
\begin{tikzpicture}[scale=0.6]
   \node at (0,0) (test2) {\wider{}};
   \node at (0,-4) (test4) {\sigwide{}};
\end{tikzpicture} 
\end{minipage}
\hfill
\vrule
\begin{minipage}{0.48\linewidth}
\centering{(iv)}  \\
\begin{tikzpicture}[scale=0.6]
   \node at (0,0) (test3) {\surround{}};
   \node at (0,-4) (test4) {\sigsurround{}};
\end{tikzpicture} \\
\end{minipage}
\begin{minipage}{0.48\linewidth}
\end{minipage}
\caption{Schematic representation of (i) localised source (ii) Multi-LS source with reverberation (RSW) (iii)  Multi-LS distributed source with angular width $\alpha$  (ESW) (iv) Surround source (LEV). Here, $\sigma^2$ represents directional signal strength}  
\label{fig_multiple_sources}
\end{figure}

\vspace*{-2mm}
\section{Binaural analysis of spatialised sources} \label{sec_wnoise}
\begin{figure*}[!t]
\centering
\begin{minipage}{0.05\linewidth}
\rotatebox{90}{\footnotesize{ITD Model}}
\end{minipage}
\begin{minipage}{0.28\linewidth}
\centering{\footnotesize{(i) Localised}}
\includegraphics[width=1\textwidth]{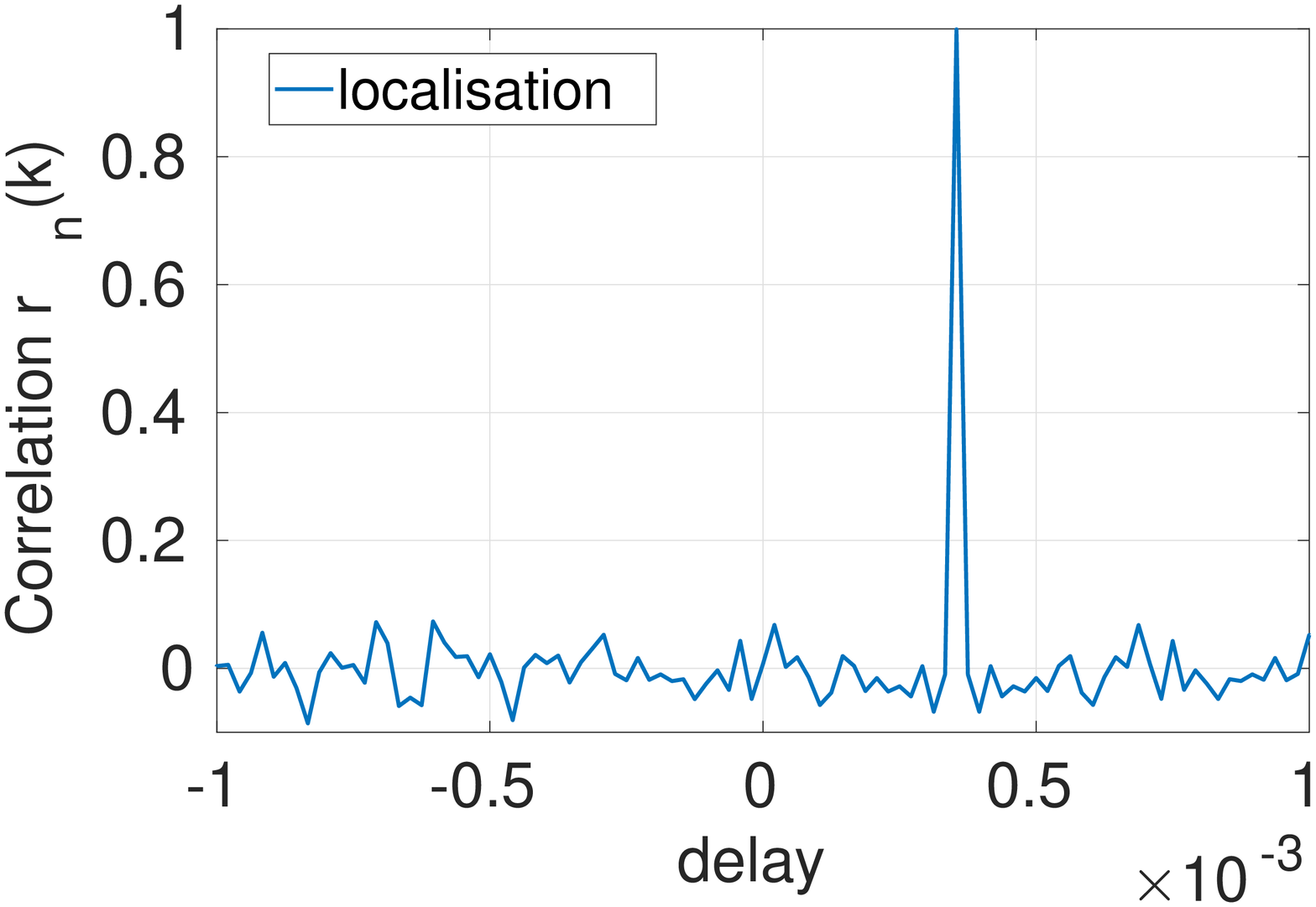}  
\centering{(a)}
\end{minipage}
\hspace{1mm}
\begin{minipage}{0.28\linewidth}
\centering{\footnotesize{(ii) Reverberation}}
\includegraphics[width=1\textwidth]{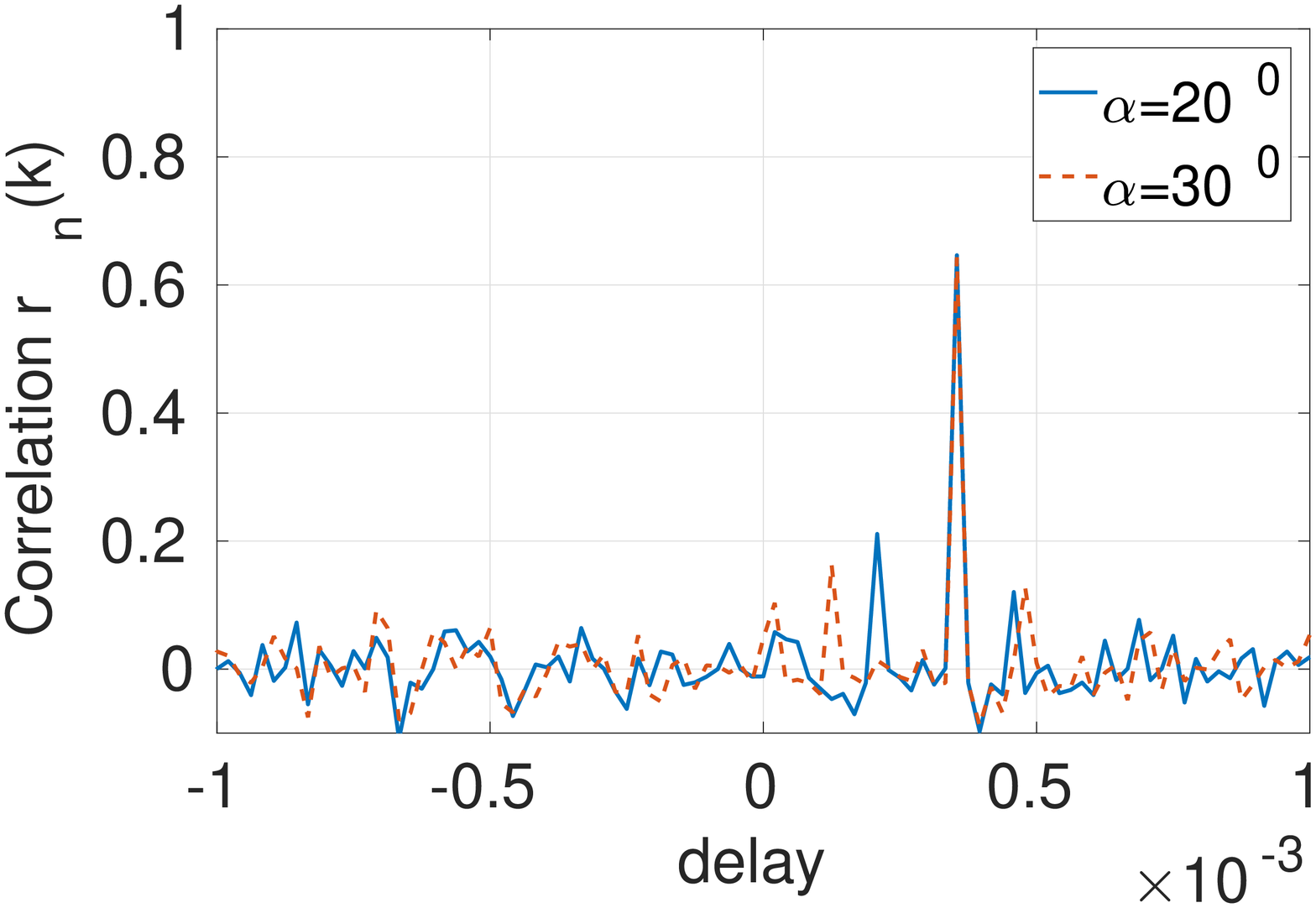}  
\centering{(b)}
\end{minipage}
\hspace{1mm}
\begin{minipage}{0.28\linewidth}
\centering{\footnotesize{(iii) Ensemble}}
\includegraphics[width=1\textwidth]{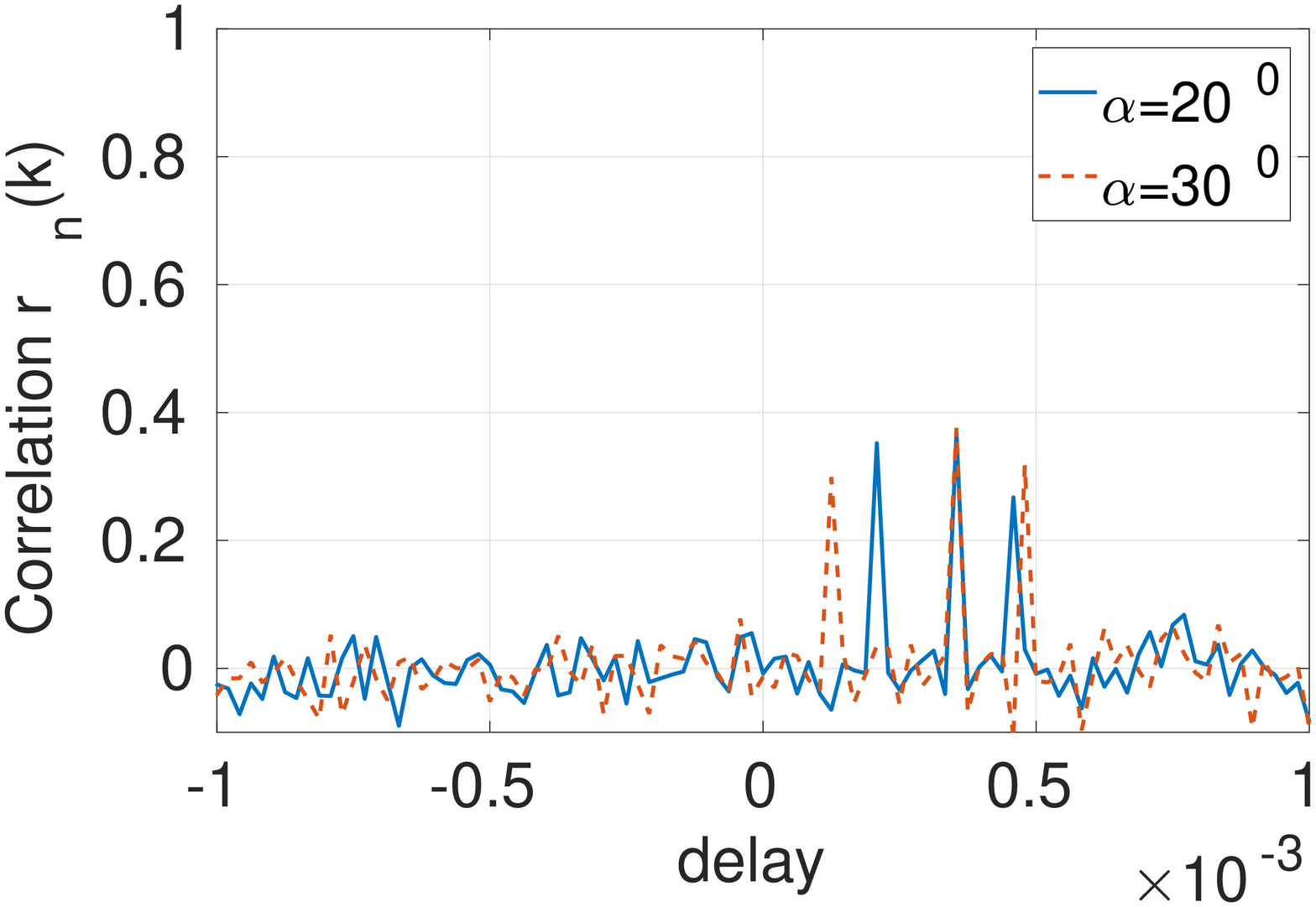}  
\centering{(c)}
\end{minipage}
\hspace{1mm}
\qquad
\begin{minipage}{0.05\linewidth}
\rotatebox{90}{\footnotesize{HRIR Model}}
\end{minipage}
\begin{minipage}{0.28\linewidth}
\includegraphics[width=1\textwidth]{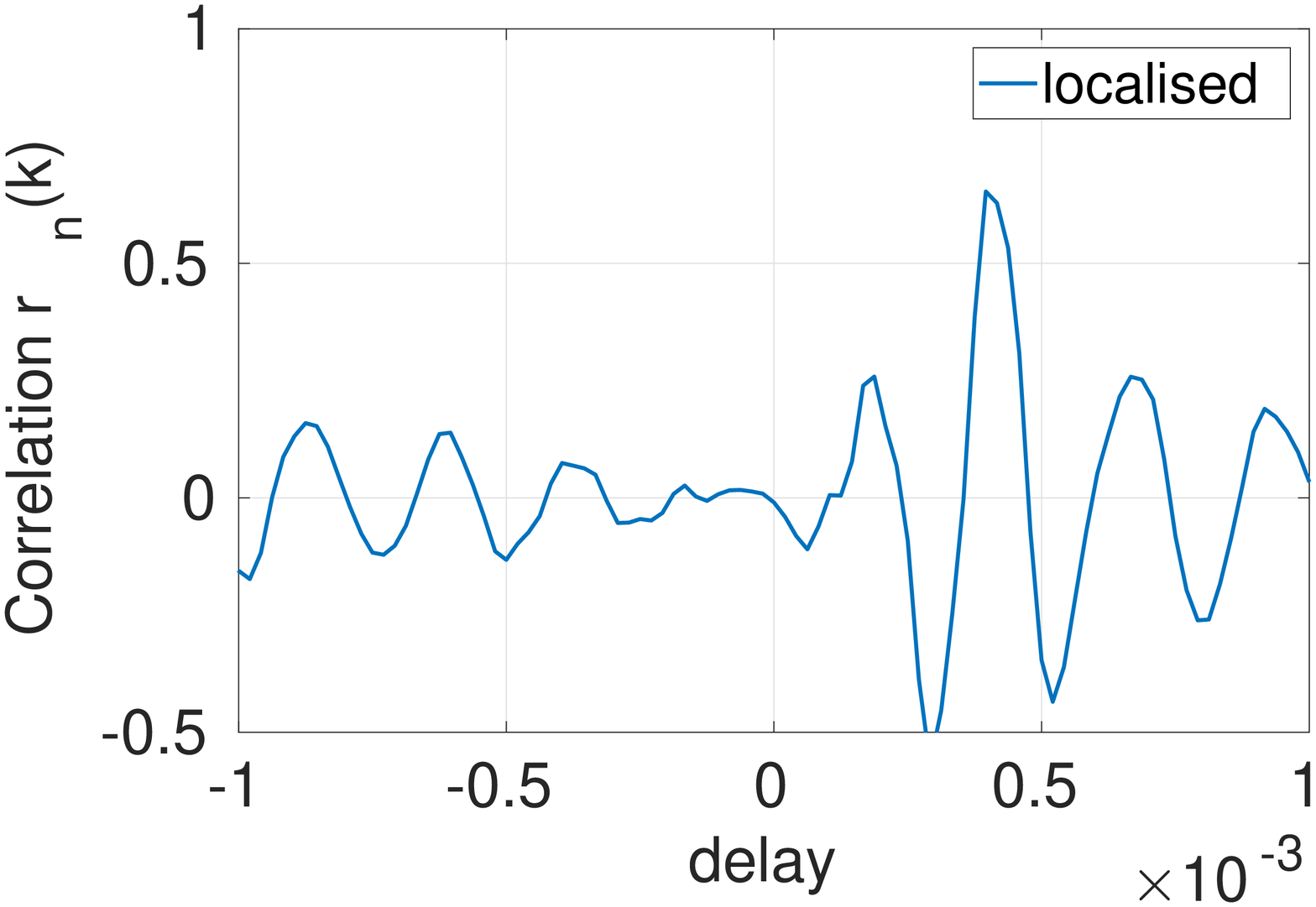}  
\centering{(d)}
\end{minipage}
\hspace{1mm}
\begin{minipage}{0.28\linewidth}
\includegraphics[width=1\textwidth]{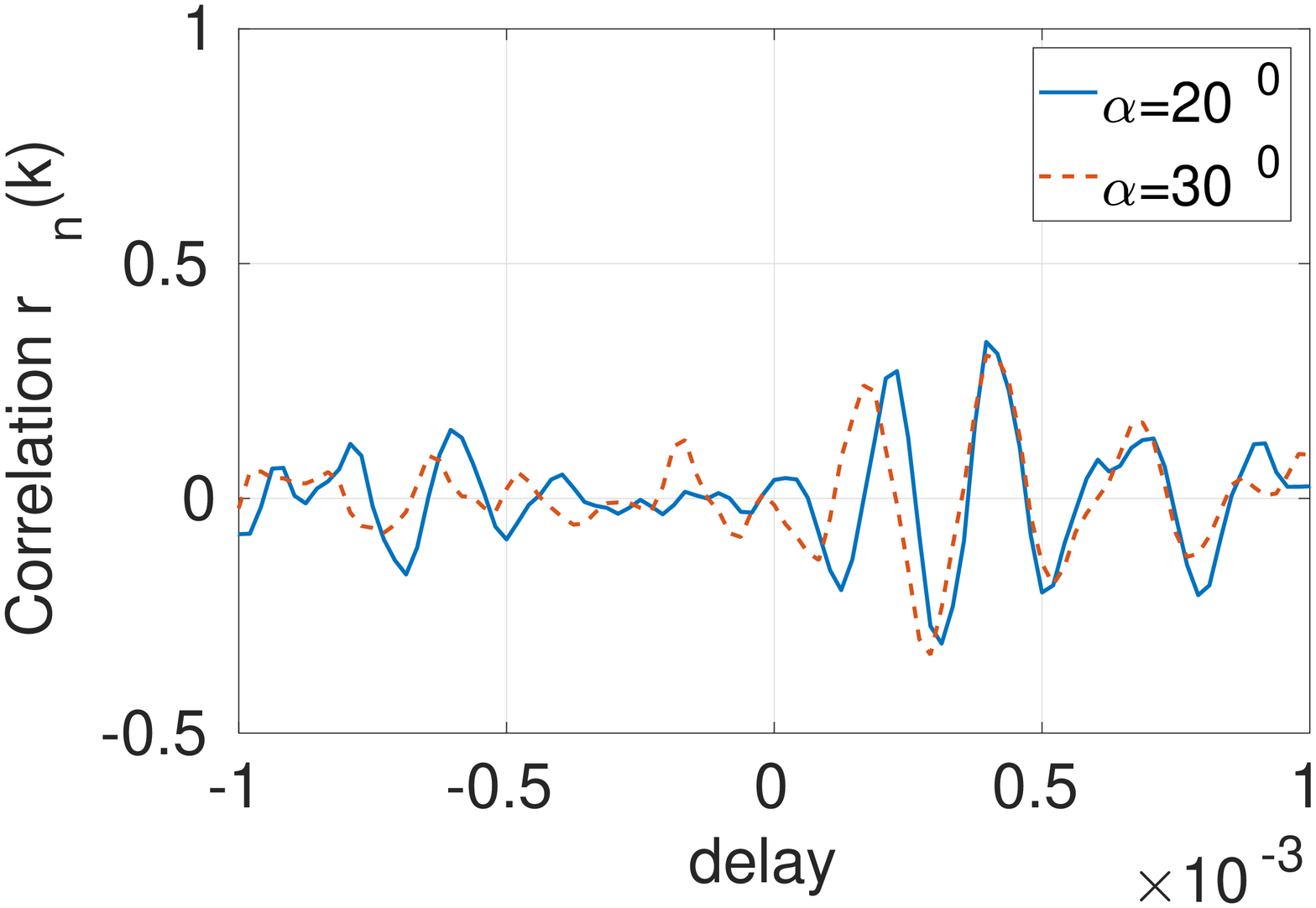}  
\centering{(e)}
\end{minipage}
\hspace{1mm}
\begin{minipage}{0.28\linewidth}
\includegraphics[width=1\textwidth]{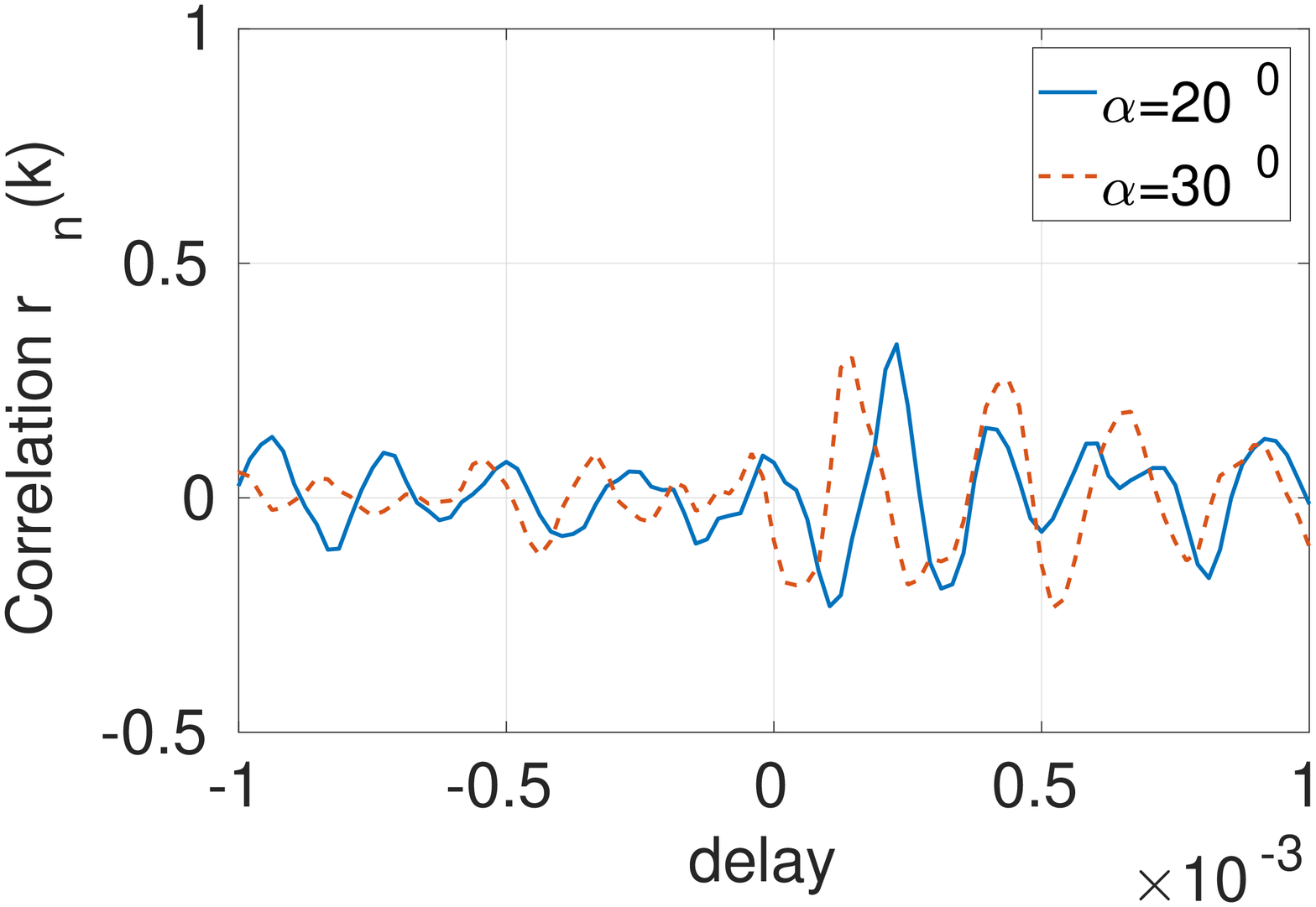}  
\centering{(f)}
\end{minipage}
\hspace{1mm}
\caption{\footnotesize{ (Color Online) Comparison of normalized cross-correlation function $r_n(k)$ for localised, reverberant and ESW cases: Simulated source with (i) Energy compact localised source at $45^0$ (ii) Dispersed energy with reverberation about at $45^0\pm20^0$ and $45^0\pm30^0$  (iii) More dispersed energy ESW with multiple sources about $45^0\pm20^0$ and $45^0\pm30^0$.  }}
\label{fig_Lsources}
\end{figure*}
\vspace*{-2mm}
Consider a multi-loudspeaker array  as shown in Fig.\ref{fig_multiple_sources}. We develop a generalised model for binaural signals for localised, RSW and ESW cases, first using uncorrelated white noise in this section and then extend the analysis to multi-component audio signals and band-pass signals (Sec.\ref{sec_corr}). Let there be L sources, each source being an uncorrelated white noise. Let $s_i(n)$ be real valued white Gaussian noise signal with normal distribution $\mathcal{N}(0,\sigma_i^2)$ played through the $i^{th}$ channel.  Let $x_r(n)$ and $x_l(n)$ be the simulated right and left channel signals at the listener position. 
\vspace*{-2mm}
\subsection{ITD model}
Let $p_i^r$ and $p_i^l$ be the path delays for right and left ears in number of signal samples and $\Delta p_i = p_i^r - p_i^l $ be the binaural path difference for the $i^{th}$ source. We can compare the binaural coherence for a localised source, reverb source and ensemble  source if they are located about the same mean position. Let  $\theta_m$ be the direction of the localised source, the dominant component of the reverb source and the mean position about which the sources are distributed in case of ensemble.  We can model a reverb source for RSW using only the early reflections\cite{morimoto1993relation}\cite{barron1981spatial}, since late reflections are considered stochastic in nature and spatially homogeneous and enveloping the listener. 
The effective right and left channel signals, simulated using ITD model are given as follows.
\begin{align}
x_r(n)  = \sum_{i=1}^{L}  s_i(n-p_i^r) \:\: ; 
x_l(n)  = \sum_{i=1}^{L}  s_i(n-p_i^l) 
\end{align}
The binaural cross-correlation for the above L no. of multiple sources is given by 
\vspace{-2mm}
\begin{align}
r(k) &= E \{ x_r(n) \cdot x_l(n-k) \}  \nonumber \\
       &= \sum_{i=1}^{L} \sigma_i^2 \delta  \big(k - (p_i^r - p_i^l)\big) 
        \nonumber  \\
        &= \sum_{i=1}^{L} \sigma_i^2 \delta  \big(k - \Delta p_i\big)  \nonumber 
\end{align}
We observe that each source is separable in correlation domain. 
\begin{itemize}[wide = 0pt]
\item {\textit {Case 1 Localised source:}}  
For a localized source, L =1 and by energy compaction property, we observe that for a given angle $\theta_m$, of all the possible spatialisations of the source about the angle $\theta_m$, an anechoic localised source is the most energy compact. Here, we assume, $\sigma_m = 1$ without loss of generality. 
\vspace*{-3mm}
\item {\textit {Case 2  Reverb source:}}
For a reverb source (RSW) with the dominant component at $\theta_m$ and attenuated components at $\theta_j, j\neq m$, the normalized IACC function is more dispersed than localised source. Here, we assume, $\sigma_m = 1$ and $\sigma_j \leq 0.5, j \neq m$ and attenuated components are distributed around $\theta_m$.
\vspace*{-2mm}
\item {\textit {Case 3  Ensemble source:}}
Let us consider an ensemble source located with a mean position $\theta_m$ and  sources distributed at $\theta_i$. For the case of ESW, we assume, $\sigma_m = 1$ and $\sigma_j \sim 1, j \neq m$, unlike reverb components. We consider the same $\theta_i$ as we considered in the case of reverb source, so that the timbre of both cases are comparable. We can compare localized sources, RSW and ESW sources when they have  same mean location $\theta_m$ and direction of arrival of dominant and attenuated components $\theta_i$. In such a case, the difference between RSW and ESW arises because of $\sigma_j, j \neq m$.  We observe that by energy compaction property, r(k) is more dispersed in the case of ESW than that of reverb source and localized source as shown in Fig.\ref{fig_Lsources}(iii). With increase in angular distribution of sources in ESW, using energy compaction property, we can infer that wider sources have more dispersed $r(k)$. 
\vspace*{-3mm}
\end{itemize}
\vspace*{-2mm}
\subsection{Binaural HRTF model} \label{sec_rtf}
\begin{figure}[!t]
\centering  
\begin{minipage}{0.45\linewidth}
\includegraphics[width=1\textwidth]{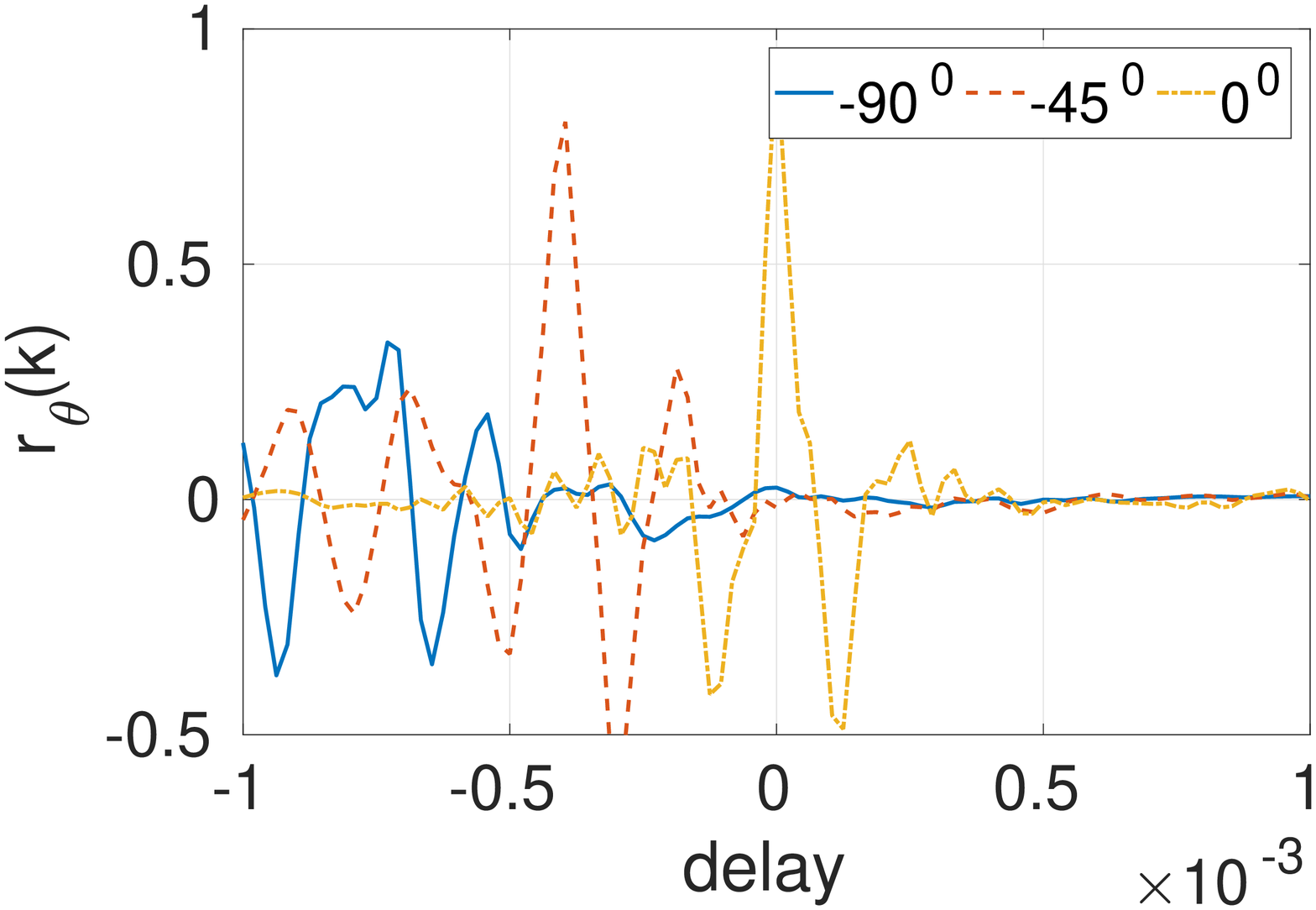}  
\centering{(a)}
\end{minipage}
\hfill
\begin{minipage}{0.45\linewidth}
\includegraphics[width=1\textwidth]{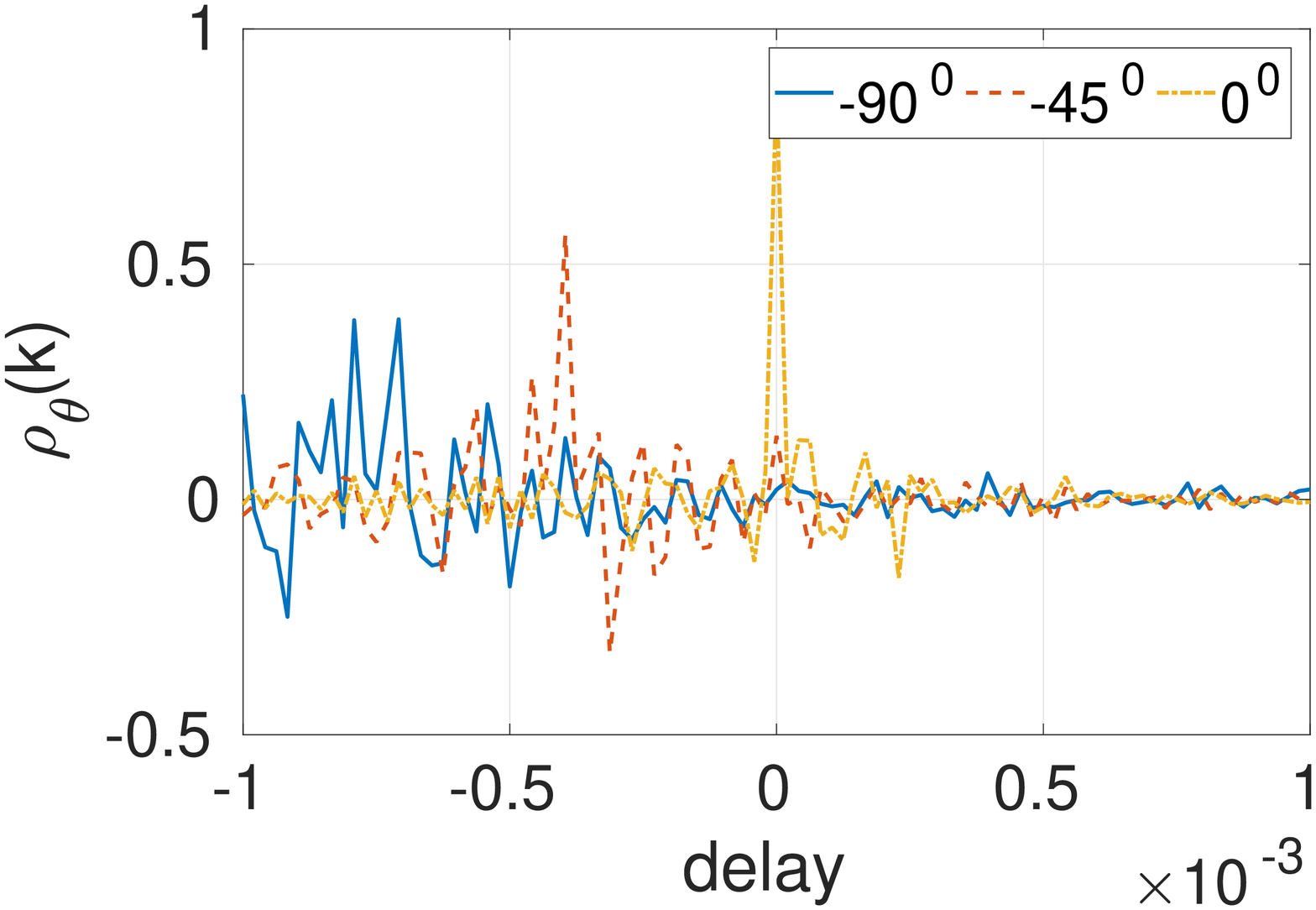}  
\centering{(b)} 
\end{minipage}
\hfill
\hfill
\caption{\footnotesize{ Difference between (a) HRIR filters  $r_\theta(k)$ and  (b) GCC-PHAT HRIR functions $\rho_{\theta}(k)$ at  source positions $0^0$, $-45^0$ and $-90^0$. We observe that the phase only HRIR functions are more compact than magnitude filters ; We also observe that functions are compact at $0^0$ in front of the listener; and sluggish at $-90^0$ and $-45^0$ corresponding to left extreme and midway between left and front of the listener.  }}
\label{fig_bcf} 
\end{figure}

Let $h_i^r(n)$ and $h_i^l(n)$ be HRIRs for right and left ear for the $i^{th}$ source. The effective right and left channel signals are given by
\vspace{-2mm}
\begin{align}
x_r(n) = \sum_{i=1}^{L} h_i^r(n) * s_i(n); \:\:\: 
x_l(n) = \sum_{i=1}^{L} h_i^l(n) * s_i(n) 
\nonumber 
\end{align}
Cross correlation of the binaural signals is given by
\begin{align}
r(k) &= E_n \{ x_r(n) \cdot x_l(n-k) \}  \nonumber  \\
             &= E_n \left\{ \sum_{i=1}^{L} h_i^r(n) * s_i(n) \cdot \sum_{j=1}^{L} h_j^l(n-k) * s_j(n-k) \right\}  \nonumber\\
             &= \sum_{i=1}^{L} \sum_{j=1}^{L} E_n  \sum_{p} h_i^r(p)s_i(n-p) \cdot \sum_{q} h_j^l(q)s_j(n-k-q)   \nonumber\\
             &= \sum_{i=1}^{L}  \sum_{j=1}^{L} \sum_{p} \sum_{q} h_i^r(p) h_j^l(q) E_n \{ s_i(n-p) s_j(n-k-q) \} \nonumber
\end{align}
\vspace{-2mm}
\begin{equation}
\text{Here,} \:\: E_n \{ s_i(n-p) s_j(n-q) \} = %
             \begin{cases}
      \sigma_i^2\delta(q-p), & \text{if }\ i=j \\
      0, & \text{if } i\neq j \nonumber
    \end{cases}
\end{equation}
\vspace*{-2mm}
\begin{align}
\therefore r(k) &= \sum_{i=1}^{L} \sum_{p} \sum_{q} h_i^r(p) h_i^l(q) \cdot \sigma_i^2\delta(k+q-p) \nonumber   \\
           &= \sum_{i=1}^{L} \sigma_i^2 \sum_{p}  h_i^r(p) h_i^l(p-k)   \\
             &\triangleq \sum_{i=1}^{L} \sigma_i^2 \cdot r_{\theta_i}(k) \label{eq_corr_filter} \\
 \text{where}\:\: &r_{\theta_i}(k) \triangleq \sum_{p}  h_{i}^r(p) h_{i}^l(p-k)  \label{eq_white_noise_corr}
\end{align}
These are the HRIR correlation filters corresponding to the direction $\theta_i$. These filters are shown in Fig.\ref{fig_bcf}(a). The centroids of the filters are separated by ITD and the maximum values are related to ILD. These filters are compact for the direction in front of the listeners and wider for lateral directions. The filters behave like directional wavelets in detecting the direction of each source. The wavelets have different resolution depending on the direction of the source as shown in Fig.\ref{fig_bcf}(a). We observe that the binaural correlation in Eq.\ref{eq_corr_filter} is a superposition of weighted HRIR correlation filters, the weights corresponding to individual source strengths ($\sigma_i^2$). The cross-correlation between left and right ear tends to be a delta-function for the ITD model. However, head related diffusion induce certain spread in $r_{\theta_i}(k)$ over the dominant delta function of the  ITD model
\vspace{-1mm}
\begin{itemize} [wide = 0pt]
 \item For the case of localised source, L=1 and we get the signature of correlation of HRIR corresponding to the location of the source as correlation function $r(k)$.
 \vspace*{-2mm}
\item For the case of a reverb source, there will be a single dominant peak and remaining L-1 no. of attenuated peak components in $r(k)$. Here, a combination of ILD and $\sigma_i^2$ component will dominate the correlation function r(k). With increase in reverberation, the auto-terms of the attenuated components also diffuse the correlation function. In the energy normalized IACC function, by increasing the reverberation, the peak value decreases and the spread in the delay domain increases. Thus,  the use of only the max value of IACC function is a over simplified measure of signal spread.  
\vspace*{-3mm}
\item For the case of  a distributed source (ensemble), L corresponds to the number of sources of nearly equal energy. Here, $\sigma_i^2$ corresponding to all L sources dominate the correlation function. Hence, the peak of IACC function will no longer be a  reliable measure for ESW, but the spatial distribution of the sources is! Hence, we try to develop a reliable measure of ESW, objectively. 
\end{itemize}
\vspace*{-2mm}

\section{Measure of ESW} \label{sec_measure}
Consider multiple distributed sources of uncorrelated noise as discussed in Sec.\ref{sec_wnoise}. We define binaural short-term correlation function and HRIR correlation function as defined in Sec.\ref{sec_rtf}. 
\begin{align}
r(k) &= \sum_{n=0}^{N-1} \{ x_r(n) \cdot x_l(n-k) \}   \\
r_{\theta}(k) &= \sum_{n=0}^{N-1}  h_{\theta}^r(n) h_{\theta}^l(n-k) 
\end{align}
Motivated by Eq.\ref{eq_corr_filter}, which includes the effect of both binaural cross-correlation as well as known HRIR, we can define a spatial correlation function as
\begin{align}
 C(\theta) \triangleq \sum_{k} r(k)\cdot r_{\theta}(k) \label{eq_sc}
\end{align}
Here, $C(\theta)$ is a spatial transform, projecting binaural correlation onto the bases of HRIR correlation. The spatial correlation function for white noise source is shown in Fig.\ref{fig_mag_phase}(a). We observe that the peaks are not clear and the magnitude effects do interfere in the interpretation of this  function and hence we define a GCC-PHAT correlation function and HRIR GCC-PHAT function as follows
\vspace*{-2mm}
\subsection{Phase-only spatial correlation function (POSC)}
For the case of a localised source with reverberation, GCC-PHAT based phase only correlation estimation provides a better estimate of localisation \cite{knapp1976generalized}. Analogous to this approach, we consider phase-only correlation functions for estimating the distribution of multiple sources.  We can define the  binaural GCC-PHAT and HRIR GCC-PHAT functions as follows.
\vspace*{-2mm}
\begin{align}
\rho(k) \xleftrightarrow{\mathscr{F}} \frac{ X_r(\omega) \cdot X_l^*(\omega)}{|X_r(\omega)||X_l(\omega)|}  \label{eq_gcc_phat} \\
\rho_{\theta}(k) \xleftrightarrow{\mathscr{F}} \frac{ H^r_{\theta}(\omega) \cdot H^{l*}_{{\theta}}(\omega)}{|H^r_{\theta}(\omega)||H^{l}_{\theta}(\omega)|}  \label{eq_gcc_phat_rtf}
\end{align}
While the magnitude inclusive functions $r(k)$ and $r_\theta(k)$ are compact filters, $\rho(k)$ and $\rho_\theta(k)$ are non-compact functions, which include higher order non-linearities also. 
\subsection*{Properties of HRTF GCC-PHAT  functions $\rho_{\theta}(k)$}
Simulated HRIR-GCC-PHAT functions $\rho_{\theta}(k)$, corresponding to measured HRIR at different directions are shown in Fig.\ref{fig_bcf}(b). It is observed that the centroid of the function within 1ms varies with the direction of the sources. The HRIR-GCC-PHAT functions have interesting properties.  While magnitude based $r_{\theta}(k)$ filter is energy compact, phase based $\rho_{\theta}k)$ has non-linearities beyond 1ms. Similarly to HRIR magnitude inclusive correlation filters, the HRIR-GCC-PHAT functions are also close to delta-function in the frontal direction around $0^0$, as shown in \ref{fig_bcf}(b) ;  The functions are sluggish near the  lateral and extreme left or right as shown in \ref{fig_bcf}(b) in the duration less than 1ms.    

\subsection*{Spatial transform - POSC function:}
\begin{figure}[t!]
\centering  
\begin{minipage}{0.45\linewidth}
\includegraphics[width=1\textwidth]{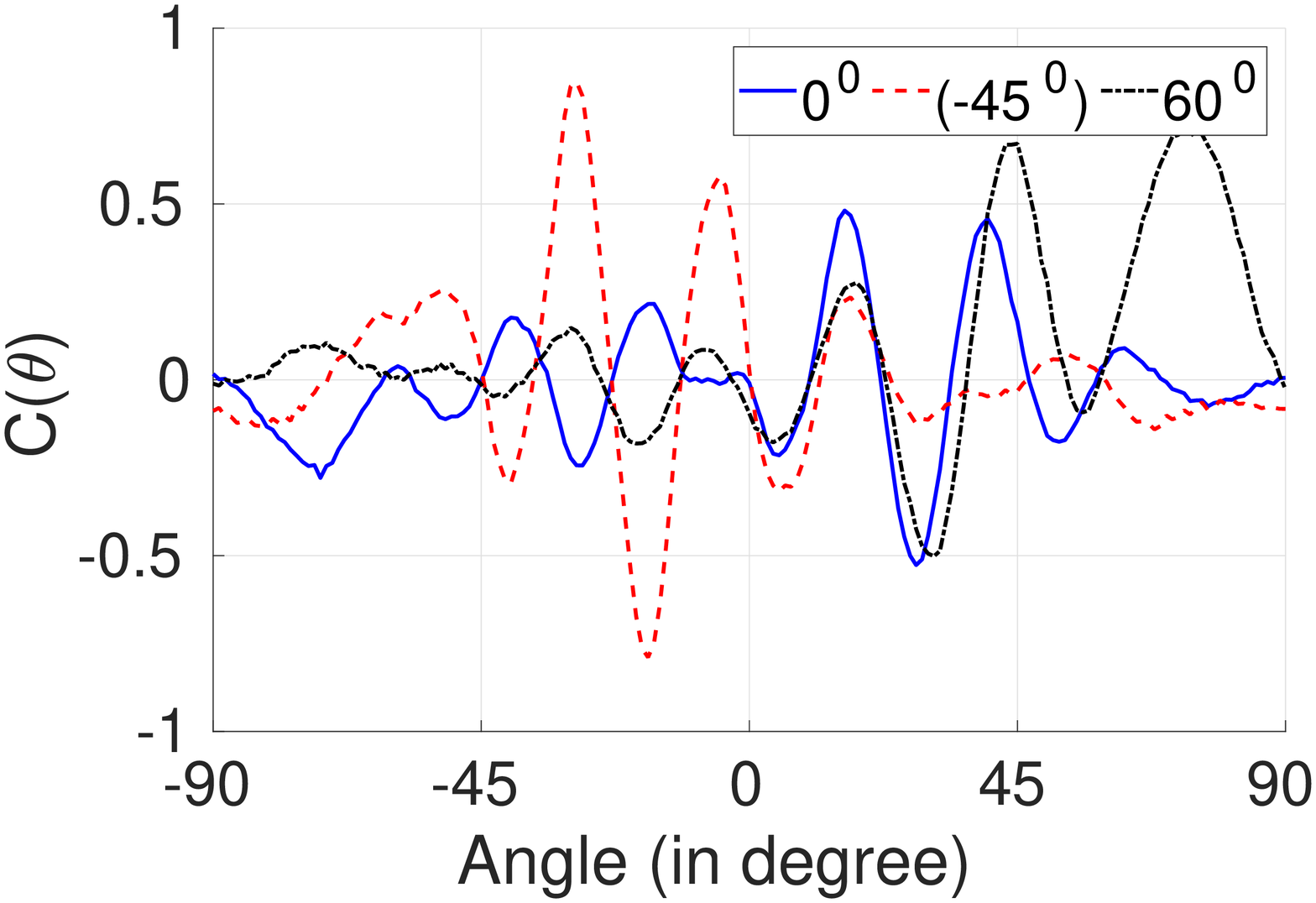}  
\centering{(a)}h
\end{minipage}
\hfill
\begin{minipage}{0.45\linewidth}
\includegraphics[width=1\textwidth]{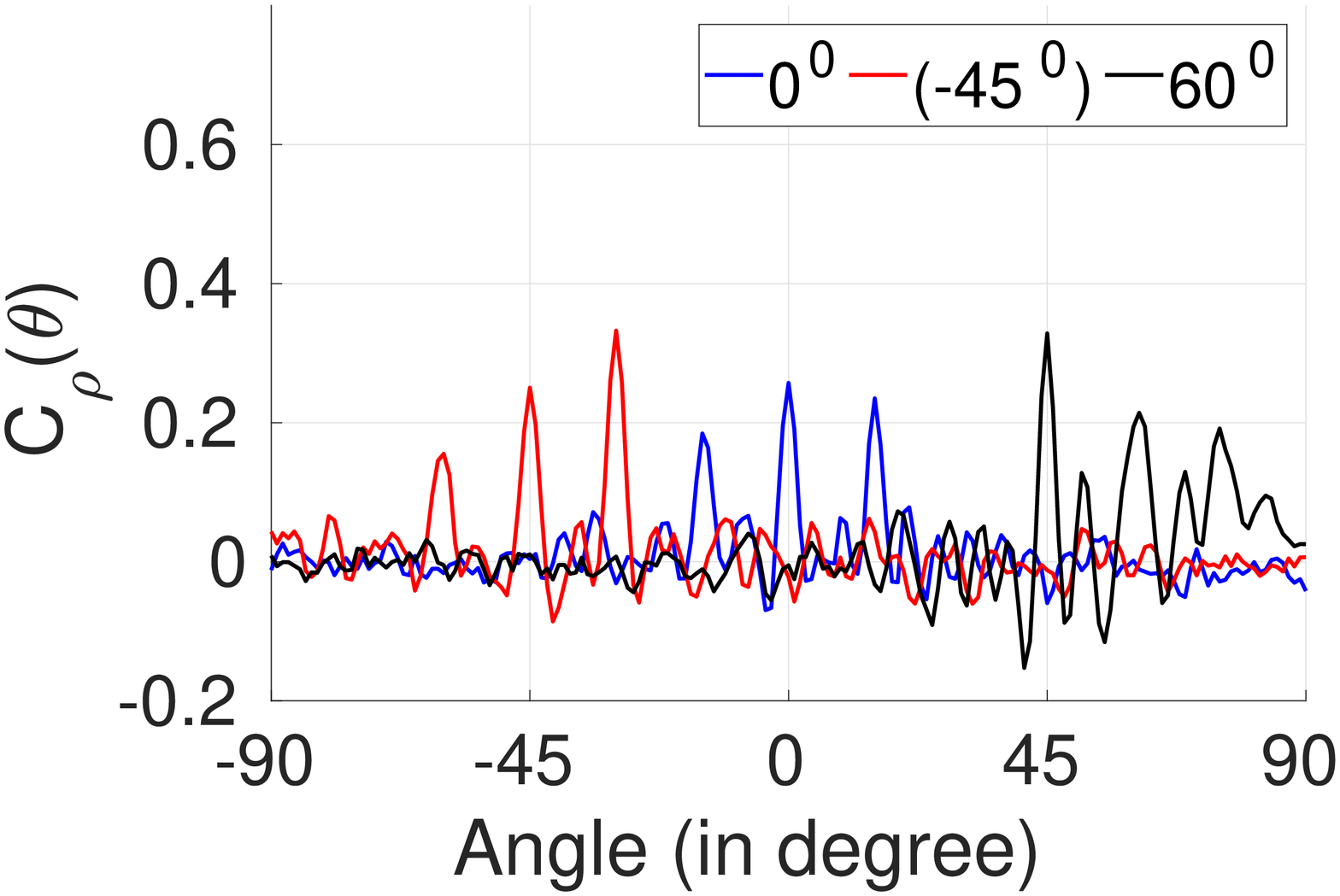}  
\centering{(b)} 
\end{minipage}
\hfill
\caption{\footnotesize{ (Color online)   Difference between (a) magnitude inclusive spatial correlation $C(\theta)$ $C(\theta)$
 and (b) phase only spatial correlation  function $C_\rho(\theta)$ for simulated wide sources of 3 distributed sources separated by $15^0$ placed about (i) $-45^0$ (red) (ii) $0^0$ (black) (iii) $60^0$ (blue) of the listener. We observe better localisation for $C_\rho(\theta)$ }}
\label{fig_mag_phase} 
\vspace*{-5mm}
\end{figure}

We define the phase-only spatial correlation (POSC) function as 
\begin{align}
 C_{\rho}(\theta) \triangleq \sum_{k} \rho(k)\cdot \rho_{\theta} (k) \label{eq_posc}
\end{align}
Here, we develop a spatial transformation by projecting phase only correlation function onto phase only HRIR correlation functional bases. The two spatial correlation functions of Eq.\ref{eq_sc} and Eq.\ref{eq_posc} , $C(\theta)$  and $C_{\rho}(\theta)$ respectively are shown in Fig.\ref{fig_mag_phase} for the case of white noise. We observe that phase only spatial correlation function  $C_{\rho}(\theta)$  helps to  estimate the location of multiple sources better than the magnitude based correlation function $C(\theta)$ . Hence, we use POSC function for estimating the directional angular extent of multi-component signals also. 
\vspace*{-2mm}
\section{Binaural model of multiple correlated signals}
\label{sec_corr}
We simulate distributed sources with a physical spread, as in an orchestra or ensemble presentation using different audio signals. We induce de-correlation by randomly delaying (+ve or -ve) the mono-channel signal for each of the channels
\cite{potard2004decorrelation}\cite{kendall1995decorrelation}. %
Thus signal played through each channel is as follows, where $d_i$ are uniformly distributed random delays samples in $\mathcal{U}$(-30ms,+30ms) and $\sigma_i$ is signal scale factor for each channel . 
\begin{align}
s_i(n) &=  \sigma_i \cdot s(n-d_i) ; S_i(\omega) = \sigma_i \cdot S(\omega)e^{-i \omega d_i} \nonumber
\end{align}
Let $h_i^r(n)$ and $h_i^l(n)$ be the HRIRs for the right and left ears from the $i^{th}$ source location. The effective right and left ear signals are expressed as follows, as a superposition of all the L no. of sources
\begin{align}
x_r(n) &= \sum_{i=1}^{L} h_i^r(n) * s_i(n) = \sum_{i=1}^{L} \sigma_i  h_i^r(n) * s(n-d_i) \nonumber  \\ 
x_l(n) &= \sum_{i=1}^{L} h_i^l(n) * s_i(n) =\sum_{i=1}^{L} \sigma_i  h_i^l(n) * s(n-d_i) \nonumber 
\end{align}
\begin{figure*}[t!]
\begin{minipage}{0.3\linewidth}
(a) sax signal \\
\includegraphics[width=1\textwidth]{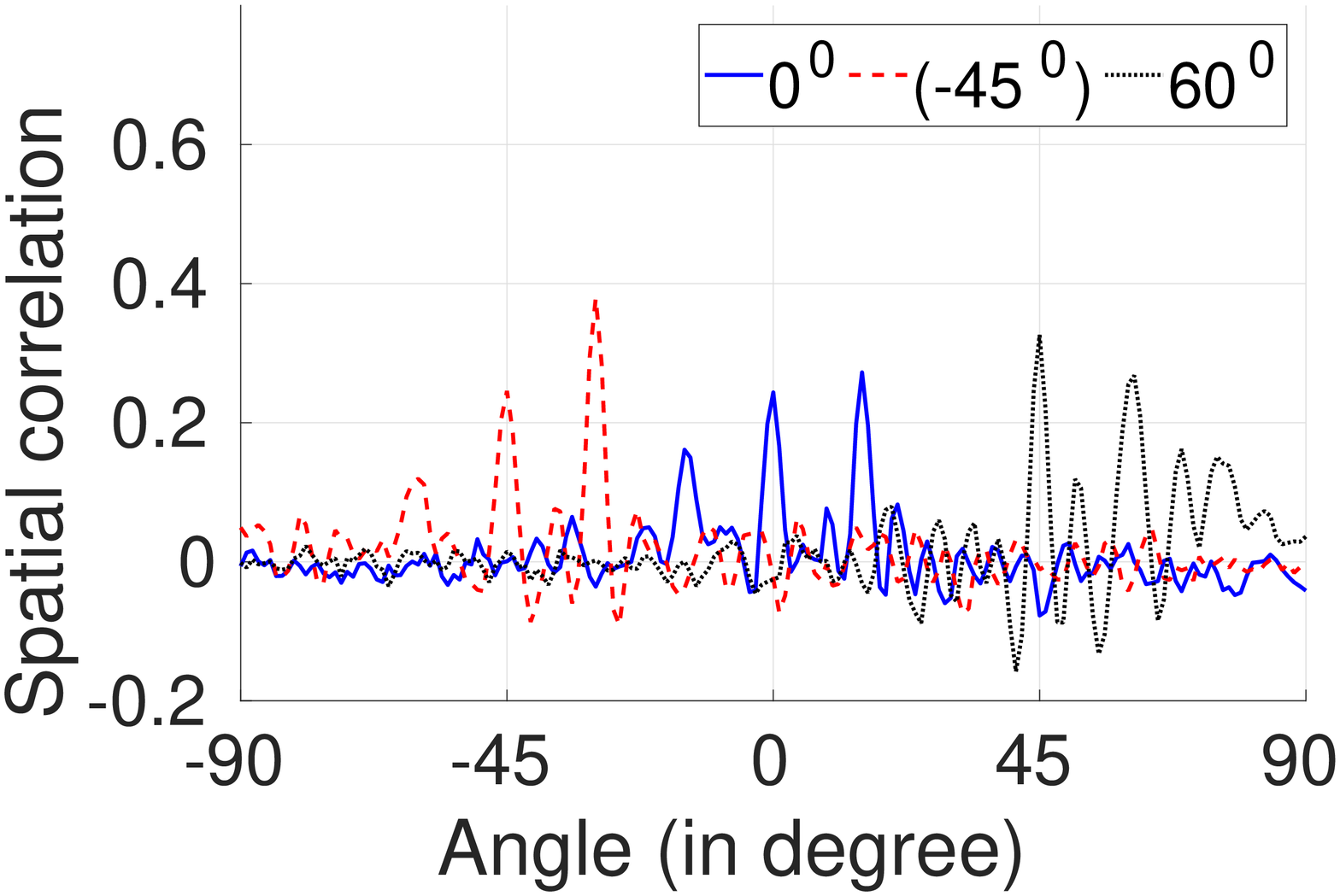}  
\end{minipage}\qquad 
\begin{minipage}{0.3\linewidth}
(b) Claves \\
\includegraphics[width=1\textwidth]{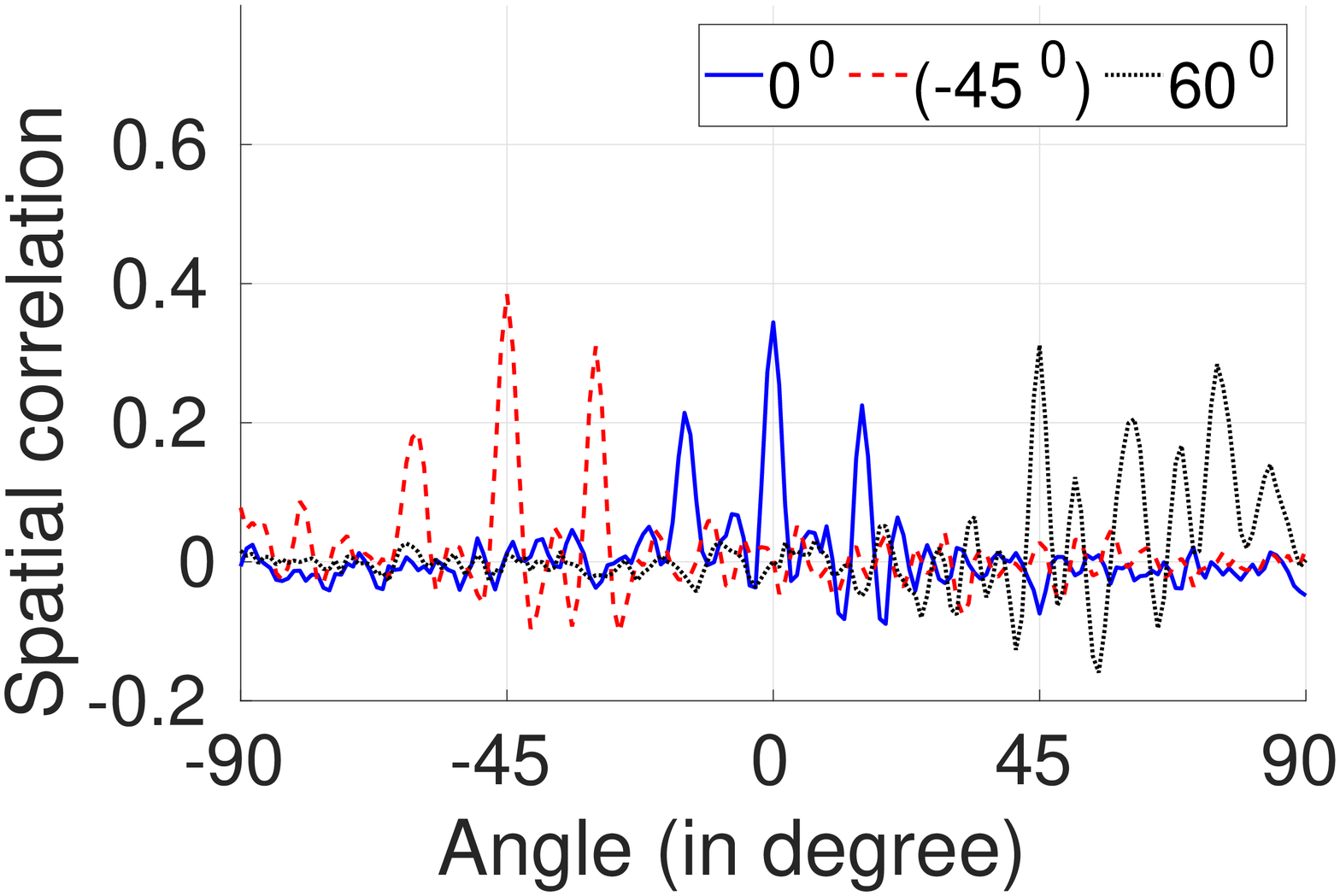}  
\end{minipage}
\begin{minipage}{0.3\linewidth}
(cd) Bandpass noise \\
\includegraphics[width=1\textwidth]{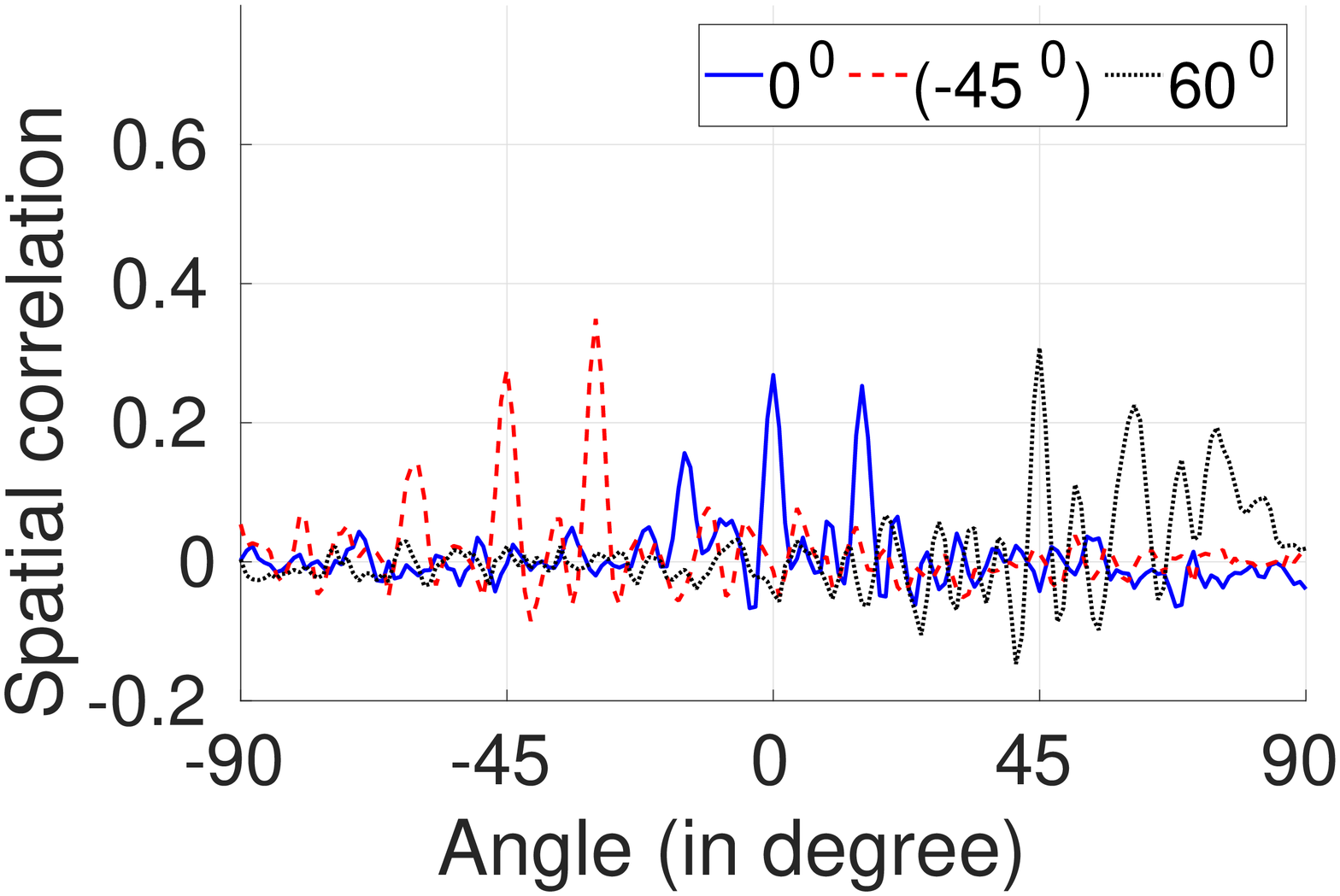}  
\end{minipage}
\caption{{(Color Online) Simulated wide sources of 3 distributed sources separated by $15^0$ placed about (i) $-45^0$ (red) (ii) $0^0$ (black) (iii) $60^0$ (blue) of the listener. Phase only Spatial correlation (POSC) for signals of different timbre (a-sax signal, b-claves, c-bandpass noise at 1k Hz). }}
\label{fig_timbre}
\end{figure*}
The corresponding Fourier transforms are 
\begin{align}
 X_r(\omega) 
	      &= S(\omega) \sum_{i=1}^{L} \sigma_i \cdot H_i^r(\omega)e^{-j\omega d_i} \nonumber  \\
 X_l(\omega) 
	     &= S(\omega) \sum_{i=1}^{L} \sigma_i \cdot H_i^l(\omega) e^{-j\omega d_i}   \nonumber
\end{align}
Cross-spectra of the binaural signals in the frequency domain and cross-correlation  are given by
\begin{align}
S_{rl}(\omega) = &X_r(\omega)\cdot X_l^*(\omega)  \xleftrightarrow{\mathscr{F}} r(k) \nonumber \\
= |S(\omega)|^2 &\sum_{m=1}^{L} \sum_{n=1}^{L} \sigma_m  \sigma_n  H_m^r(\omega) H_n^{l*}(\omega)   e^{-j\omega(d_m-d_n)}              
\end{align}
The auto-terms (m=n) are of low modulation frequency phase difference components and cross-terms $(m\neq n)$ are high modulation frequency phase difference components because of de-correlation terms $d_m$ and $d_n$. The lower order delay elements of $r(k)$ are predominantly because of the auto-terms  and non-linearities from the cross-terms. As the number of sources increase, the strength of non-linearities due to cross-terms also increases. 
\par 
We simulate wide sources with multi-component audio signals like saxaphone (sustained instrument), claves (transient instrument) and band-pass noise. We compute binaural GCC-PHAT using Eq.\ref{eq_gcc_phat} for these signals and also, compute the POSC function of Eq.\ref{eq_posc}. These are as shown in Fig.\ref{fig_timbre}. We observe that peaks are found at the location of each source even for in the signals of different timbre. Hence, we propose that POSC function  is a timbre-independent indicator of ESW. The angular extent can also be measured as  the difference in angle between the extreme source peaks. We observe that there are attenuated spurious peaks and also the effect of ILD. It is required to compensate for ILD and develop a POSC based function which can estimate the relative directional strengths of the sources. 

\vspace*{-3mm}
\section{Spatiogram: Effect of timbre} \label{sec_timbre} 
\begin{figure*}[!t]
\begin{minipage}{0.05\linewidth}
\end{minipage}
\begin{minipage}{0.28\linewidth}
$\alpha=0^0$
\end{minipage}
\begin{minipage}{0.28\linewidth}
$\alpha= 13^0$
\end{minipage}
\begin{minipage}{0.28\linewidth}
$\alpha= 21^0$
\end{minipage}
\\
\begin{minipage}{0.05\linewidth}
\rotatebox{90}{Sax}
\end{minipage}
\begin{minipage}{0.28\linewidth}
\includegraphics[width=1\textwidth]{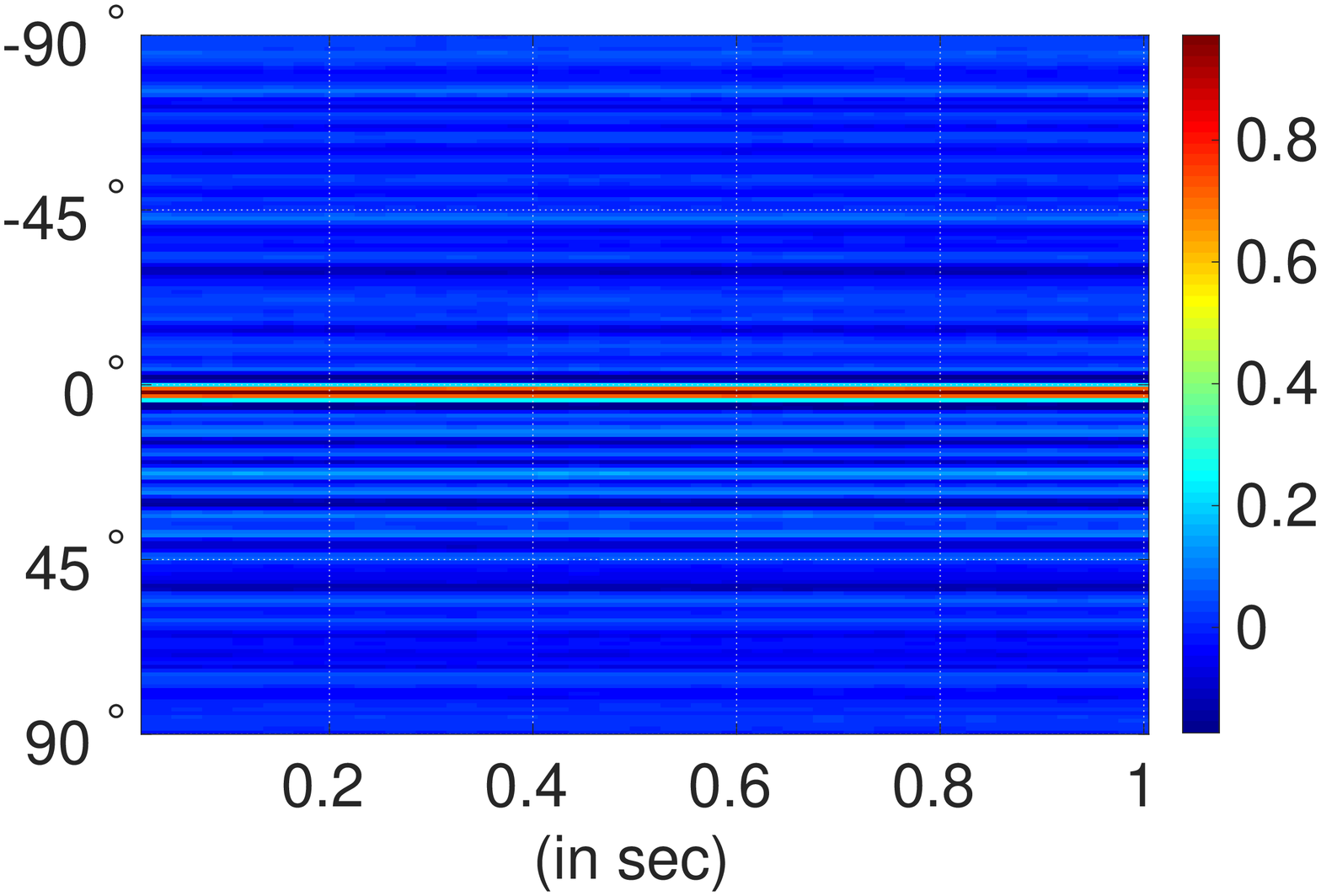}  
\end{minipage}  
\hfill
\begin{minipage}{0.28\linewidth}
\includegraphics[width=1\textwidth]{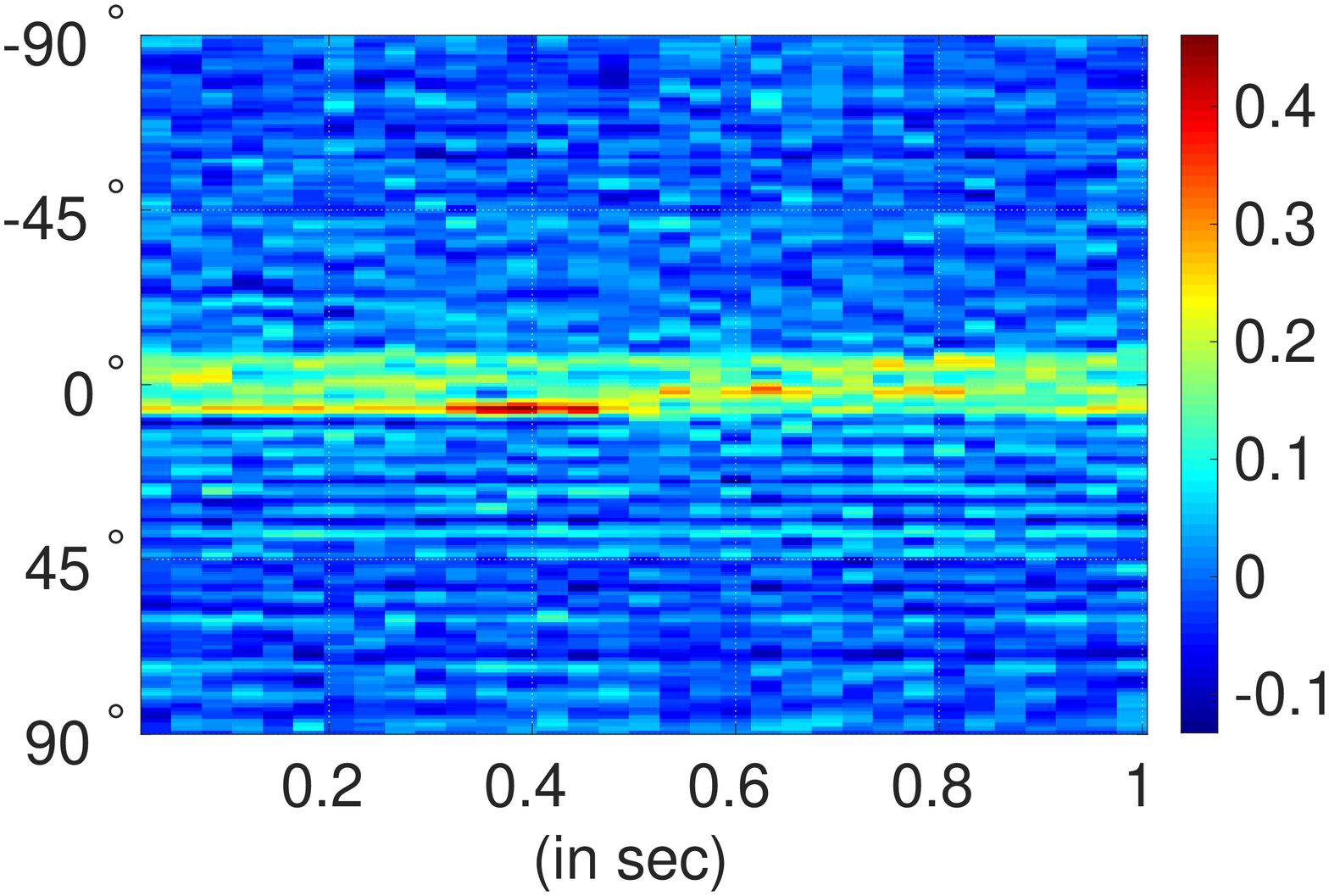}  
\end{minipage}  
\hfill
\begin{minipage}{0.28\linewidth}
\includegraphics[width=1\textwidth]{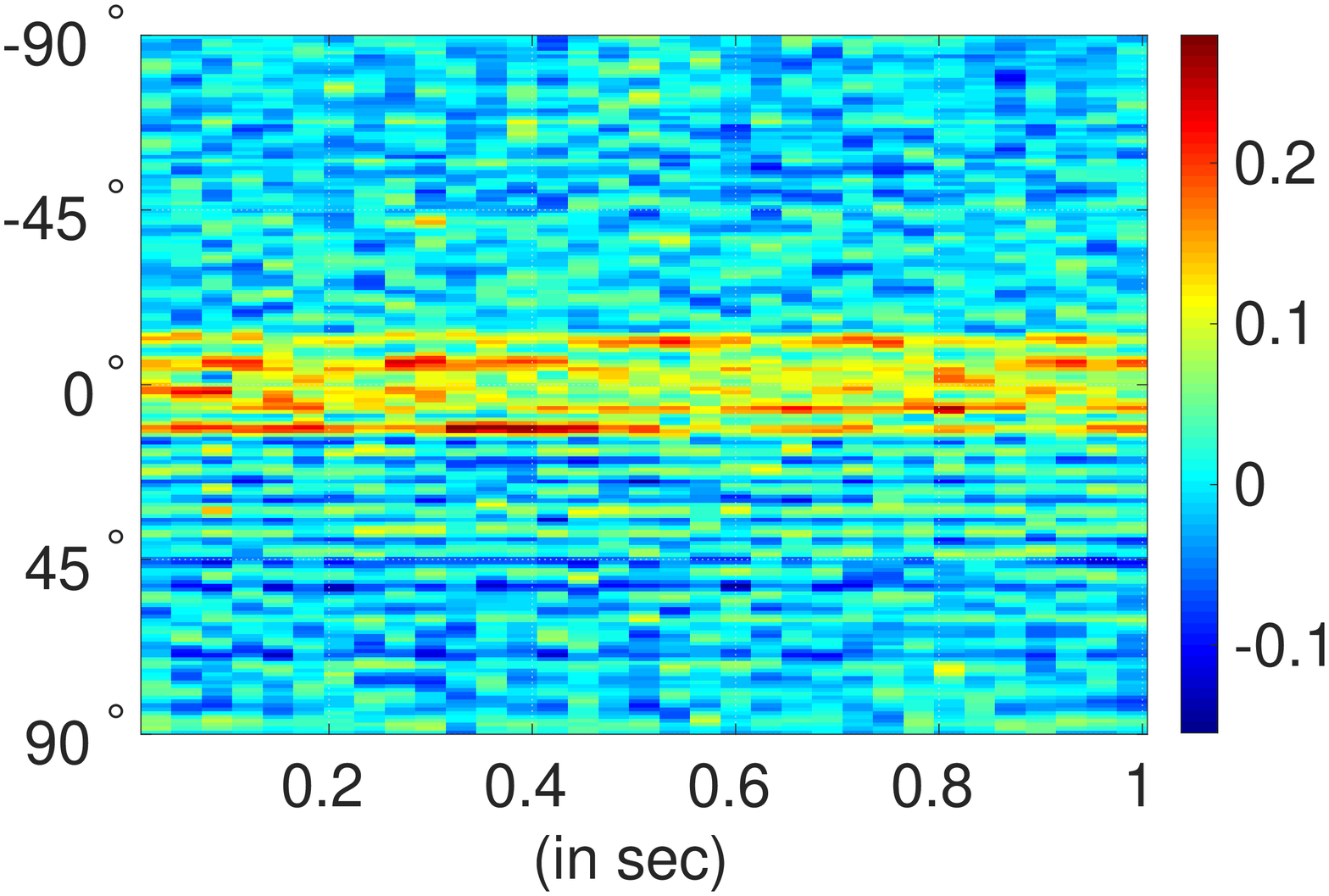}  
\end{minipage}  
\\
\begin{minipage}{0.05\linewidth}
\rotatebox{90}{Claves}
\end{minipage}
\begin{minipage}{0.28\linewidth}
\includegraphics[width=1\textwidth]{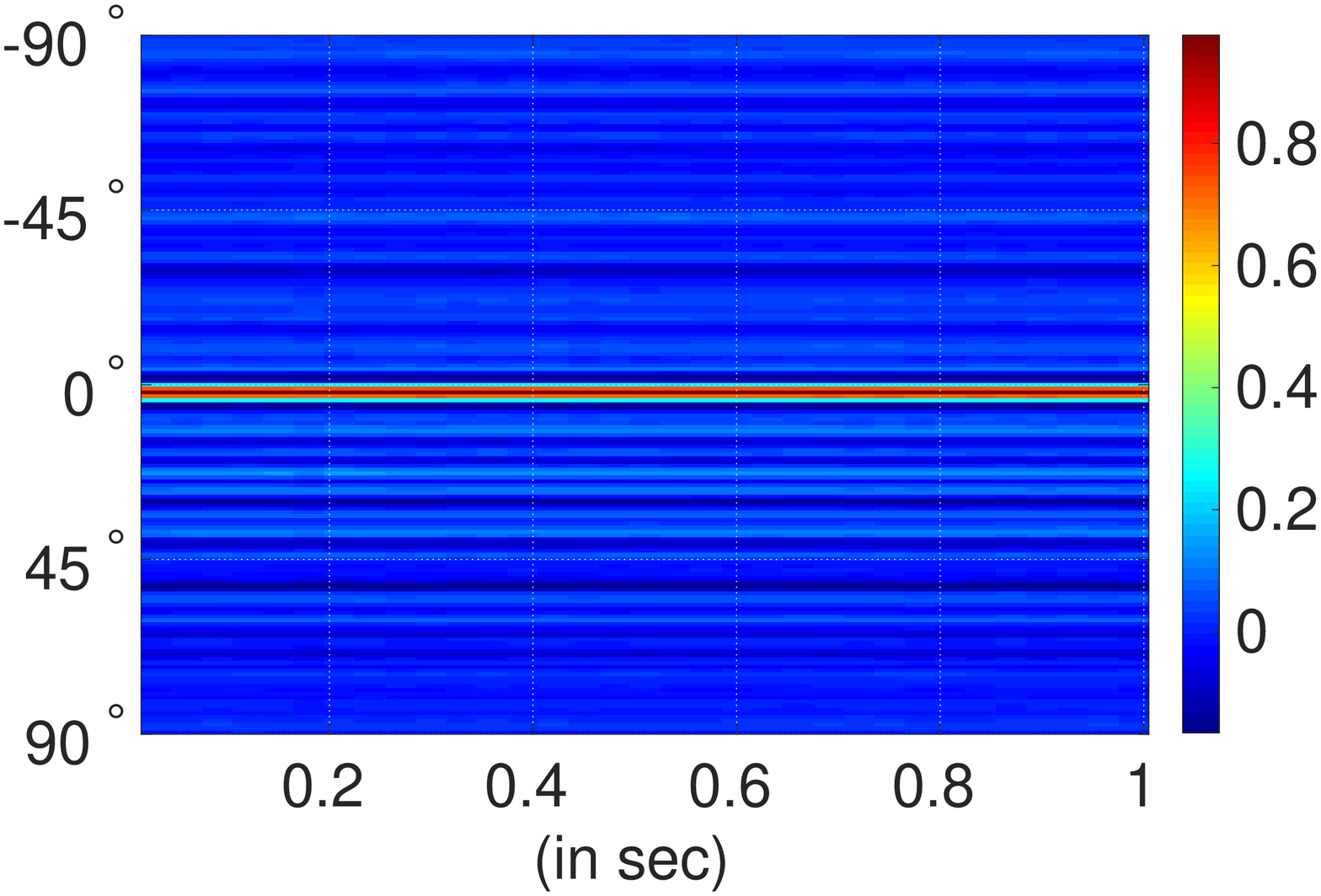}  
\end{minipage}
\hfill
\begin{minipage}{0.28\linewidth}
\includegraphics[width=1\textwidth]{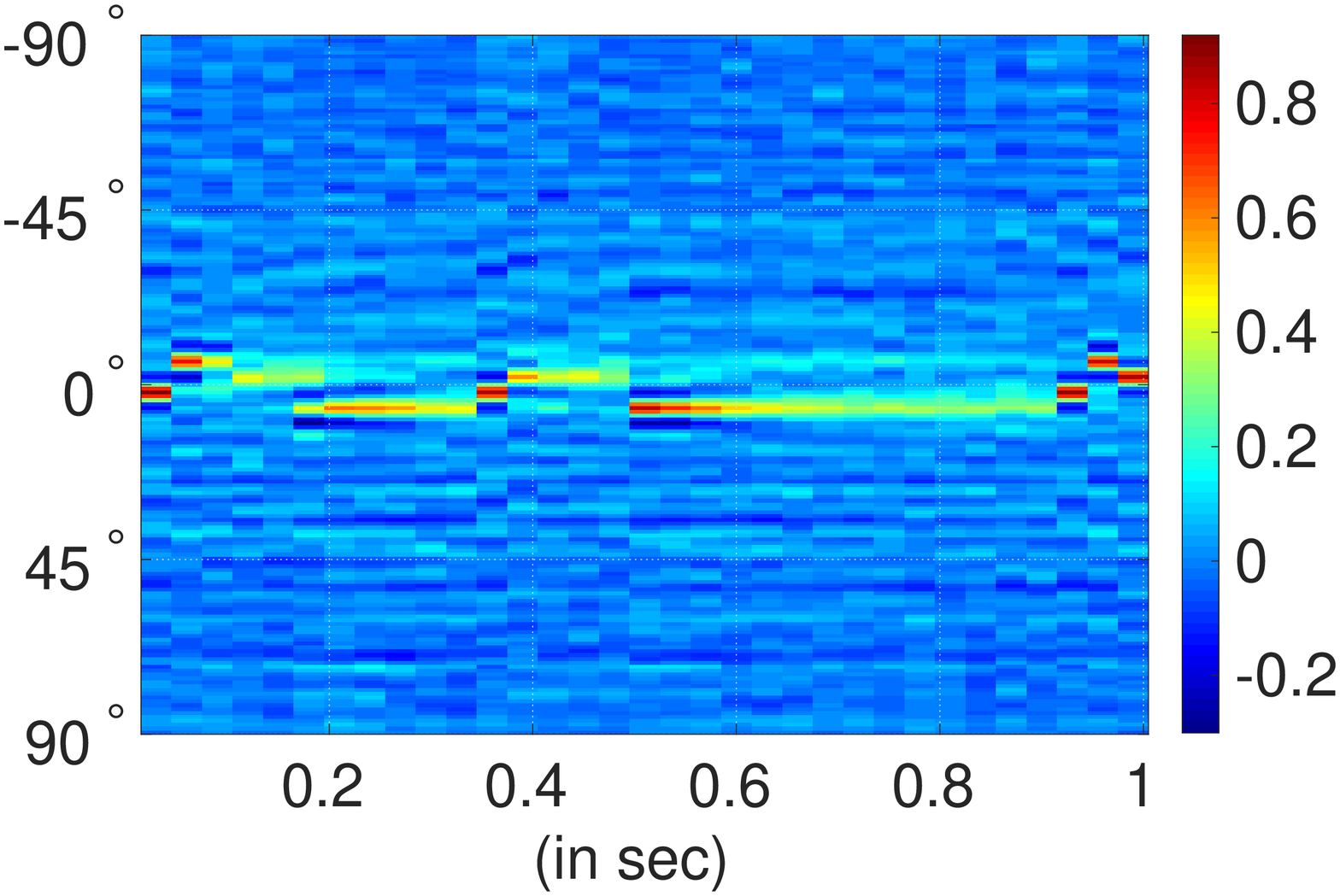}  
\end{minipage}
\hfill
\begin{minipage}{0.28\linewidth}
\includegraphics[width=1\textwidth]{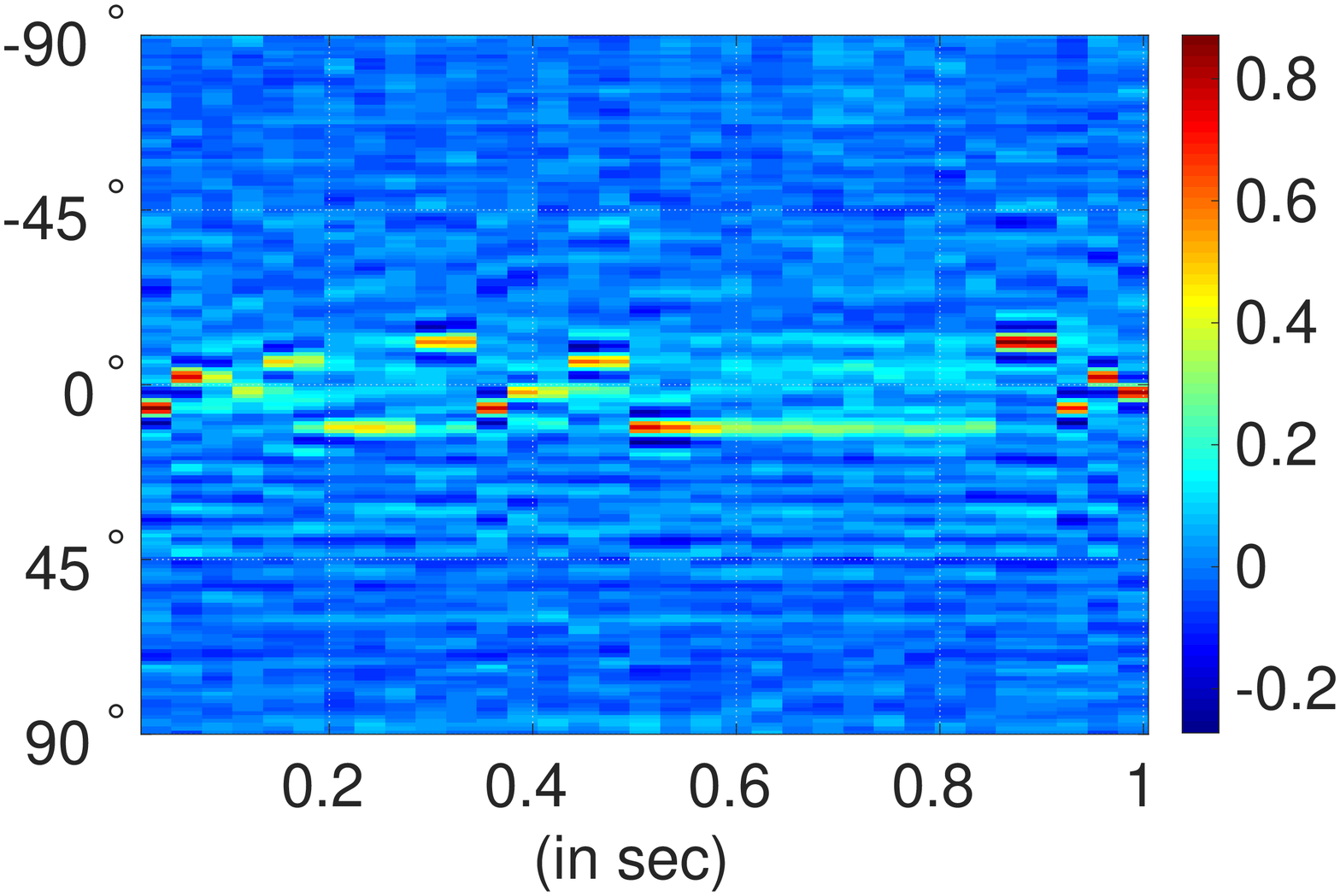}  
\end{minipage}
\caption{(Color Online) Spatiogram: Sustained (sax) and transient (claves) signals of different angular spread}
\label{fig_spatiogram}
\end{figure*}
We define a short-term time-varying phase only spatial correlation function as follows , similar to Eq.\ref{eq_sc}, but using  Eq.\ref{eq_gcc_phat} and Eq.\ref{eq_gcc_phat_rtf}, where $\rho(m,k)$ represents the binaural cross-correlation at the time window m.
\begin{align}
 C_{\rho}(m,\theta) \triangleq \sum_{k} \rho(m,k)\cdot \rho_{\theta} (k)
\end{align}
Here, $\rho_{\theta} (k)$  is  independent of time, since it is only HRIR dependent. Analogous to spectrogram, where short-term frequency variations are estimated, we estimate here the short-term spatial variations in the azimuth and define it as ``spatio-gram''. This measure can be extended to include elevation also as HRIRs are available.  In this work, we use a frame size of 80ms with $50\%$ window shift. As observed in the  time-varying spatial correlation function or the spatio-gram  shown in Fig.\ref{fig_spatiogram}, we find that for sustained signals, all the sources are active at all time instants. With increase in width around $0^0$, the dispersion in spatio-gram increases. In Fig.\ref{fig_spatiogram}, we find that for transient signals, the sources are active at different time instants because of the low energy decay regions. With increase in width around $0^0$, the dispersion in spatio-gram is not uniform as observed for sustained instruments, but more impulsive because of time-varying energy distribution in transient signals with attack and decay. 
\vspace*{-3mm}
\section{Experiment}  \label{sec_exp}
In the literature, different experimental methods have been used to explore ESW. In the approach of Pulkki et al \cite{hirvonen2006perception} \cite{santala2009resolution}, \cite{santala2011directional}, listeners are asked to locate each individual loudspeaker for the sound source. The inaccuracy in the  localization is interpreted as "integrated perception". In another approach \cite{otani2017largeness}, using a single loudspeaker and reproduced sound using headphones, the perceived largeness was rated on a horizontal and vertical perceptual scale. In the study of perceptual frequency sensitivity of IACC \cite{mason2005frequency}, de-correlated sinusoids are presented binaurally and sensitivity to IACC variations is inferred as contributing to width perception. In the present work, we render ensemble of distributed sources of given music signals, using multiple loudspeakers. We ask the listeners to rate the width of each presentation on a continuous scale using MUSHRA-like evaluation method \cite{series2014method}. Thus, we evaluate the relative source width of different timbral signals directly with  explicit reference stimuli and probe the integrated perception leading to ESW. \par
\begin{figure}[!b]
\includegraphics[width=0.8\linewidth]{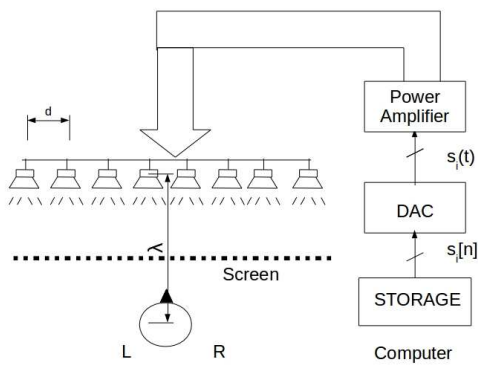}
\caption{{Scheme of  rendering a spatially wide source.$s_i[n]$ are rendered as de-correlated multi-channel signals}}
\label{fig_ls_setup}
\end{figure}
\begin{figure}[b!]%
\centerline{\includegraphics[width=0.8\linewidth]{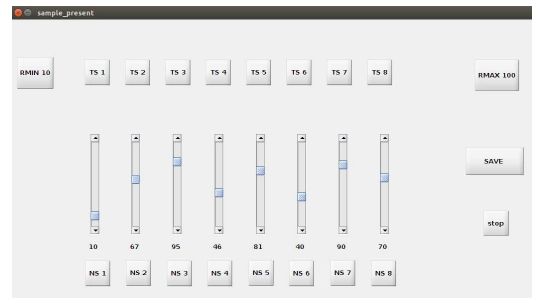}}
\caption{{GUI for the perception test of ESW }}
\label{fig_gui}
\end{figure}

The experimental setup to quantify perceptual ESW through the multi-loudspeaker  rendering is shown in Fig.\ref{fig_ls_setup}. We choose a linear array of full range frequency response loudspeakers (Boston Acoustics HQ48002P), each separated by 30 cm. A MUSHRA like listening test is developed where the test stimuli are to be compared with explicit  reference stimuli as shown in Fig.\ref{fig_gui}. More details on the experimental setup, stimuli and experimental methodology can be referred to \cite{arthi2019multi}.

\subsection{ ESW Stimuli}
\begin{table*}
\centering
\caption{ {Music instruments for ASW perception}}
\label{table_musical_instruments}
\begin{tabular}{ | p{0.7cm} | p{1.2cm} | p{1.2cm} | p{1.2cm} |p{1.2cm} |p{1.8cm} |p{1.5 cm} |p{1.5 cm} |p{1.3cm} |p{1.3 cm} |p{1.3cm} |} 
\hline
{S.No} & {Stimuli type} & \multicolumn{9}{|c|}{Instruments}\\
 \hline
1 & String & Celesta & Grand piano & Guitar & Harp & Organ & Piano & Viola  & Violin1 & Violin2 \\
2 & Vocals & Bass male & Tenor male & Alto female & Soprano female 1 & Soprano female 2 & Speech male & Speech Female & & \\
3 & Wind & Clarinet & Flute & Horn	& Oboe	& Piccolo & Sax & Trom-hbone & Trumpet & Tuba \\
4 & Percuss-ion & Bell & Castanet & Claves & Glockenspiel & Drum1 & Drum2 & Vibra-phone & Xylo &     \\
\hline
\end{tabular}
\end{table*}
We have chosen four sets of stimuli from the SQAM database as given in Table.\ref{table_musical_instruments} . 
We have used stereo recorded signal of different music instruments at a sampling frequency $f_s=48$ KHz and we  choose only the first channel of the stereo recording. The stimuli comprise of music tunes played using different individual instruments (string/ wind/ percussion and vocal singing). We have also included male and female speech as stimuli for ensemble testing. Each stimuli is approximately of  duration $\small{\sim}$15s. 
The stimuli of different instruments vary widely in their loudness range, pitch range, timbre and duration \cite{arthi2019multi}. Signal for each channel of ensemble presentation is generated using the de-correlation model \cite{potard2004decorrelation}\cite{kendall1995decorrelation} of 
\begin{align}
s_i[n] = s[n-d_i] 
\end{align}
Here $d_i\in\mathcal{U}(-30ms,+30ms)$ in number of samples. Experimentally, it is observed that for sustained signals of violin, oboe, horn, etc, different randomized inter-channel delays result in similar width perception and hence the specific choice of the values of $d_i$ is not critical.
\subsection{Listening Test}
\begin{figure*}
 \centering
 \begin{minipage}{0.4\linewidth}
  \includegraphics[width=7cm]{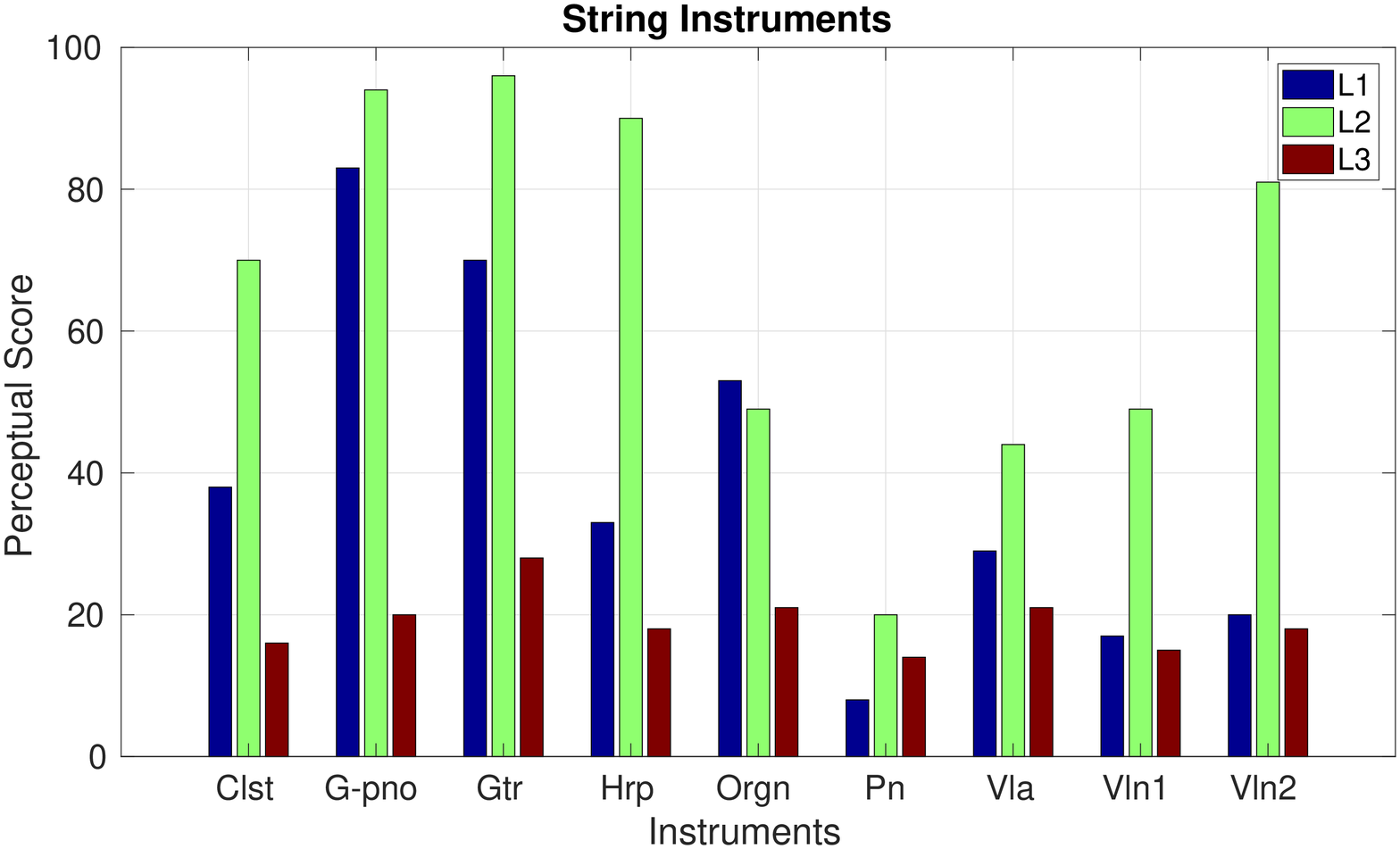}
  \end{minipage}
 \begin{minipage}{0.4\linewidth}
 \includegraphics[width=7cm]{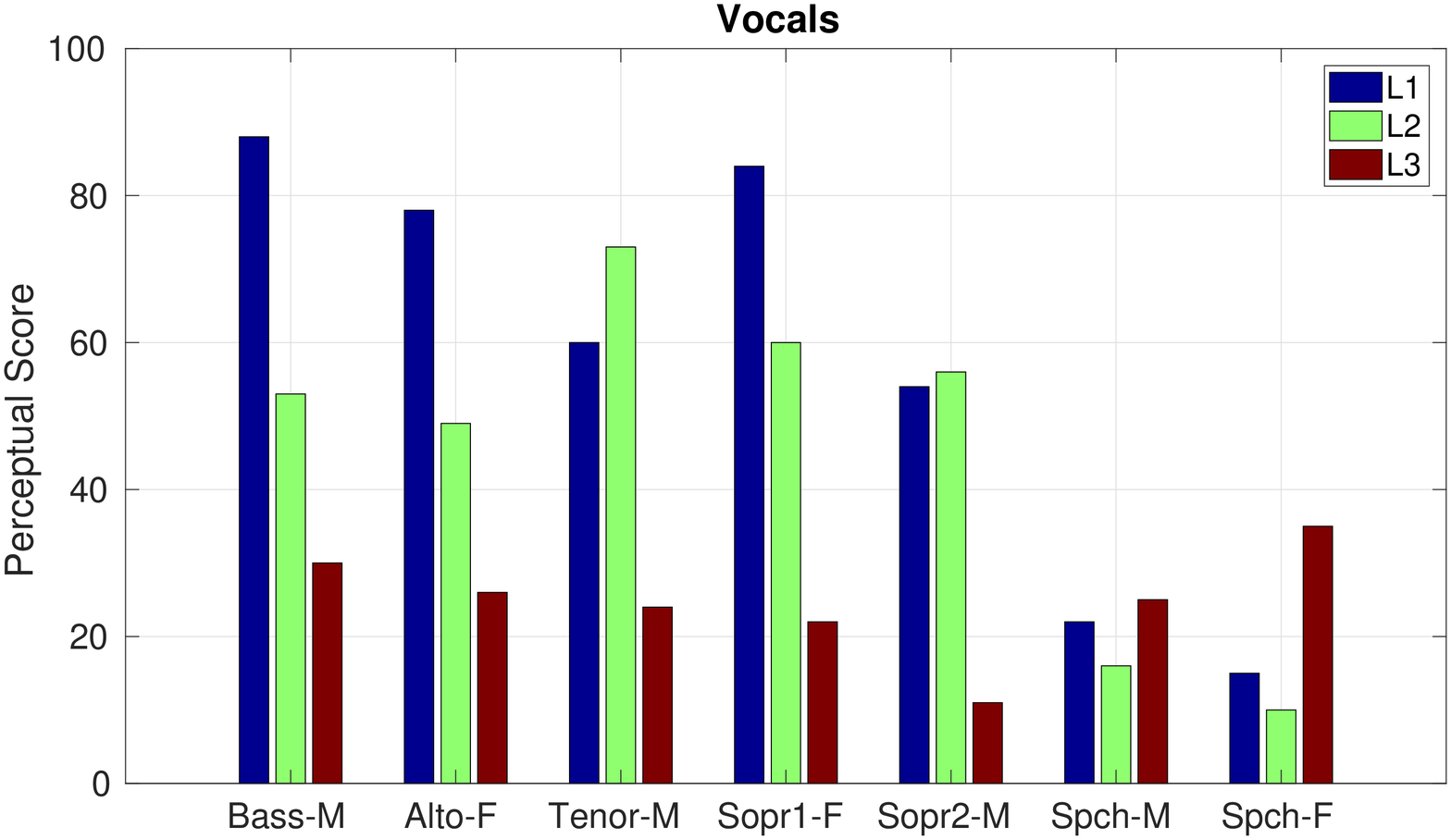}
 \end{minipage}
 \begin{minipage}{0.4\linewidth}
\includegraphics[height=4cm]{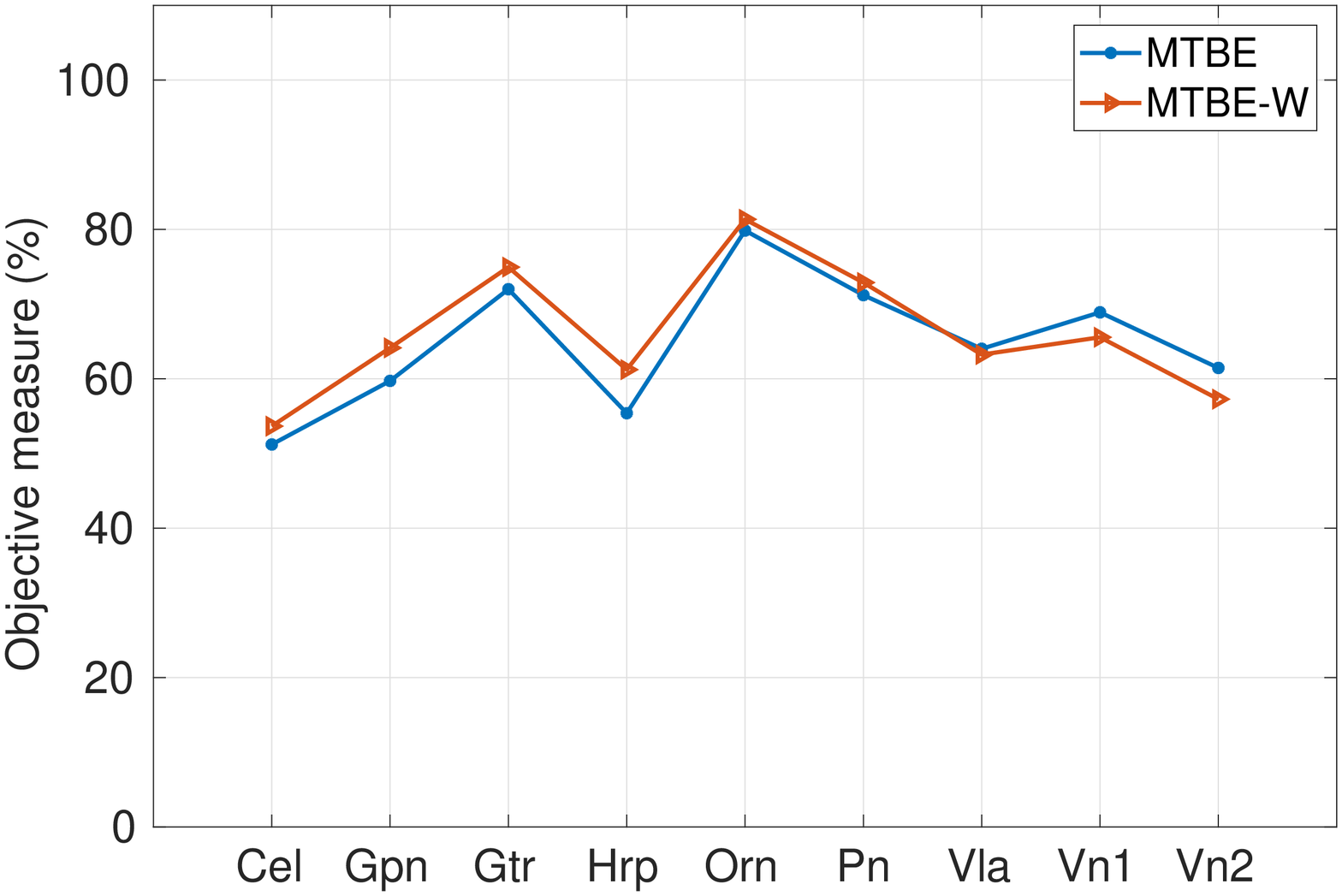}
 \end{minipage}
 \begin{minipage}{0.4\linewidth}
 \includegraphics[height=4cm]{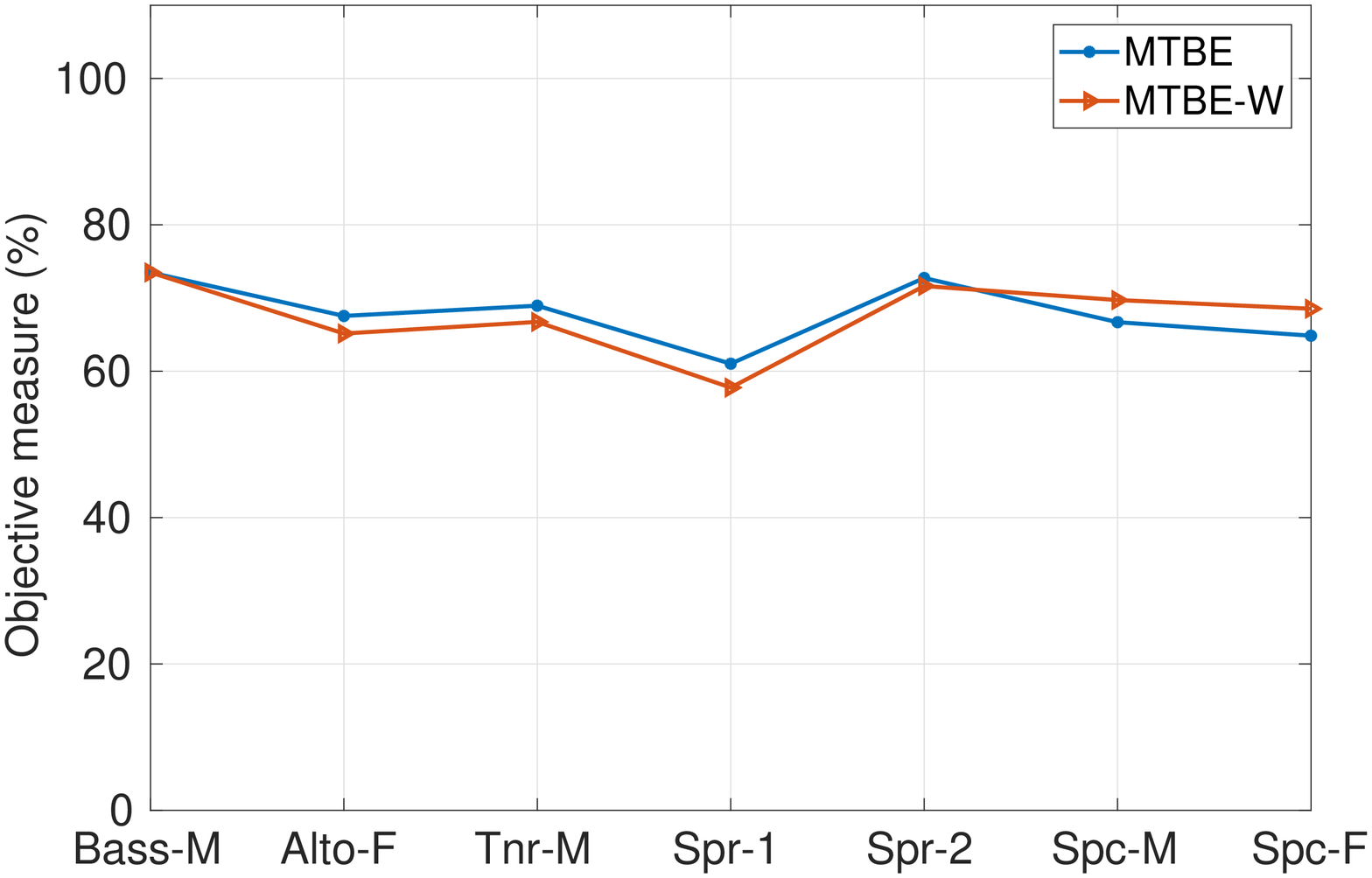}
 \end{minipage}
 \begin{minipage}{0.4\linewidth}
 \includegraphics[width=7cm]{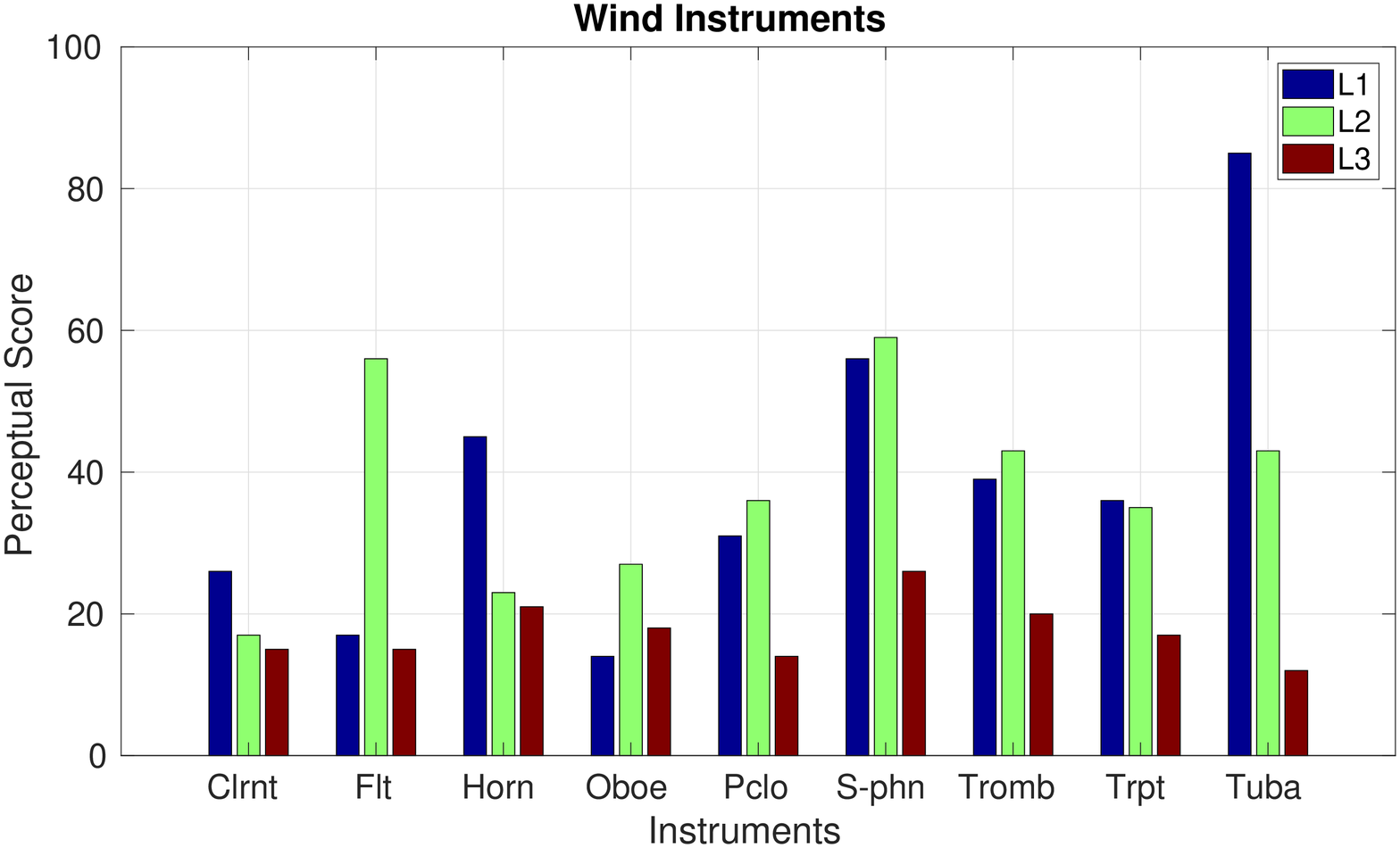}
 \end{minipage}
 \begin{minipage}{0.4\linewidth}
 \includegraphics[width=7cm]{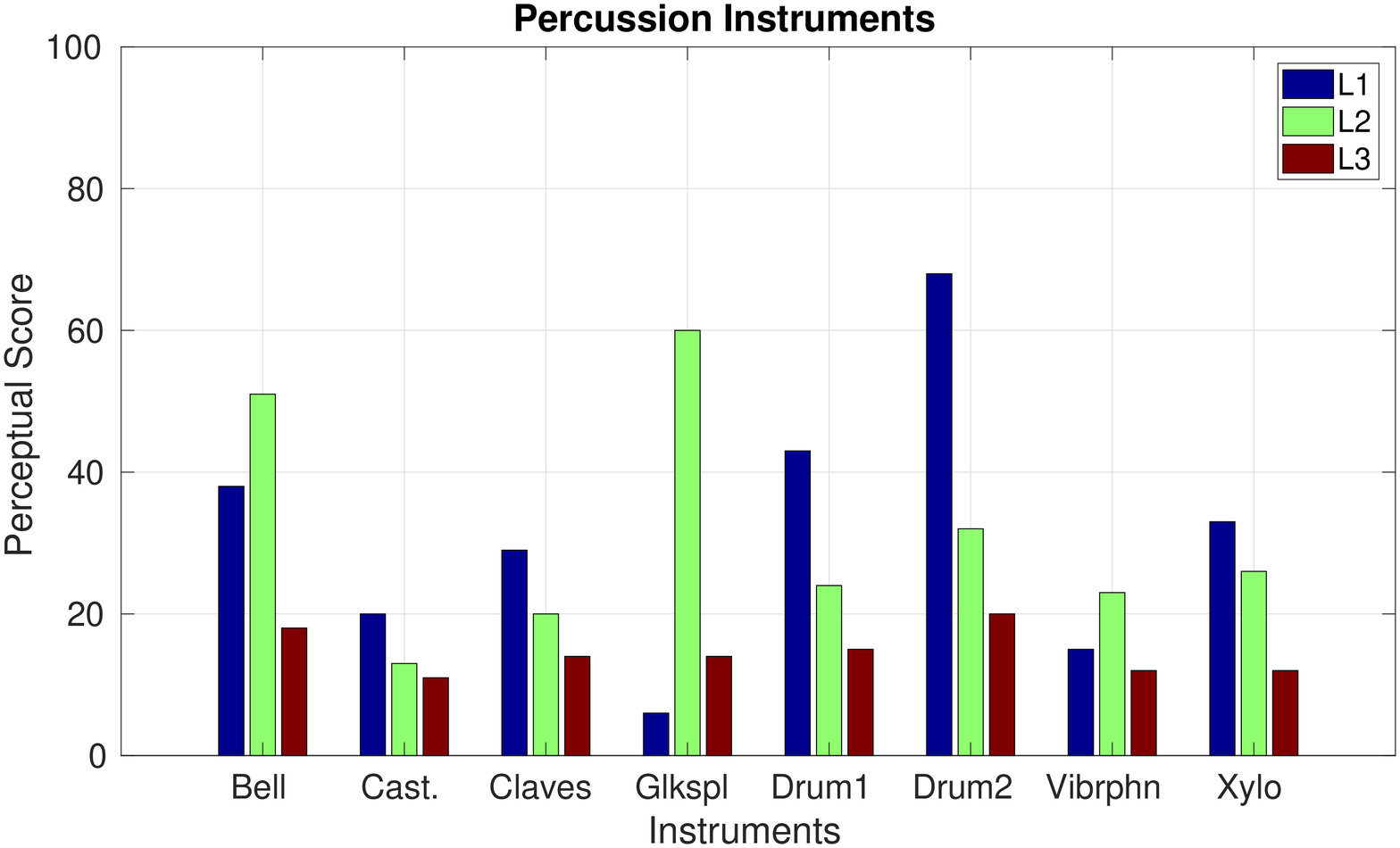}
 \end{minipage}
 \begin{minipage}{0.4\linewidth}
 \includegraphics[height=4cm]{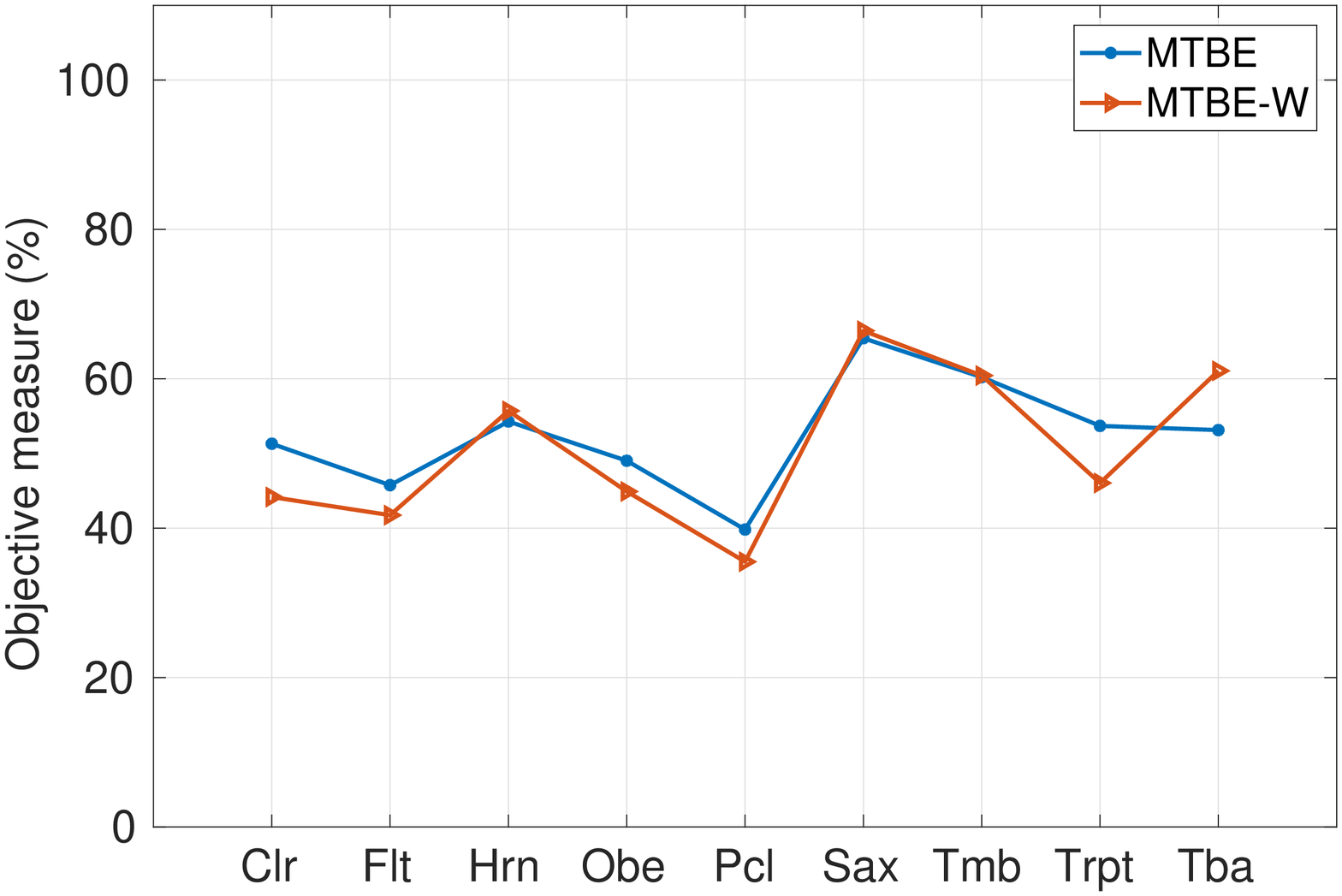}
 \end{minipage}
 \begin{minipage}{0.4\linewidth}
 \includegraphics[height=4cm]{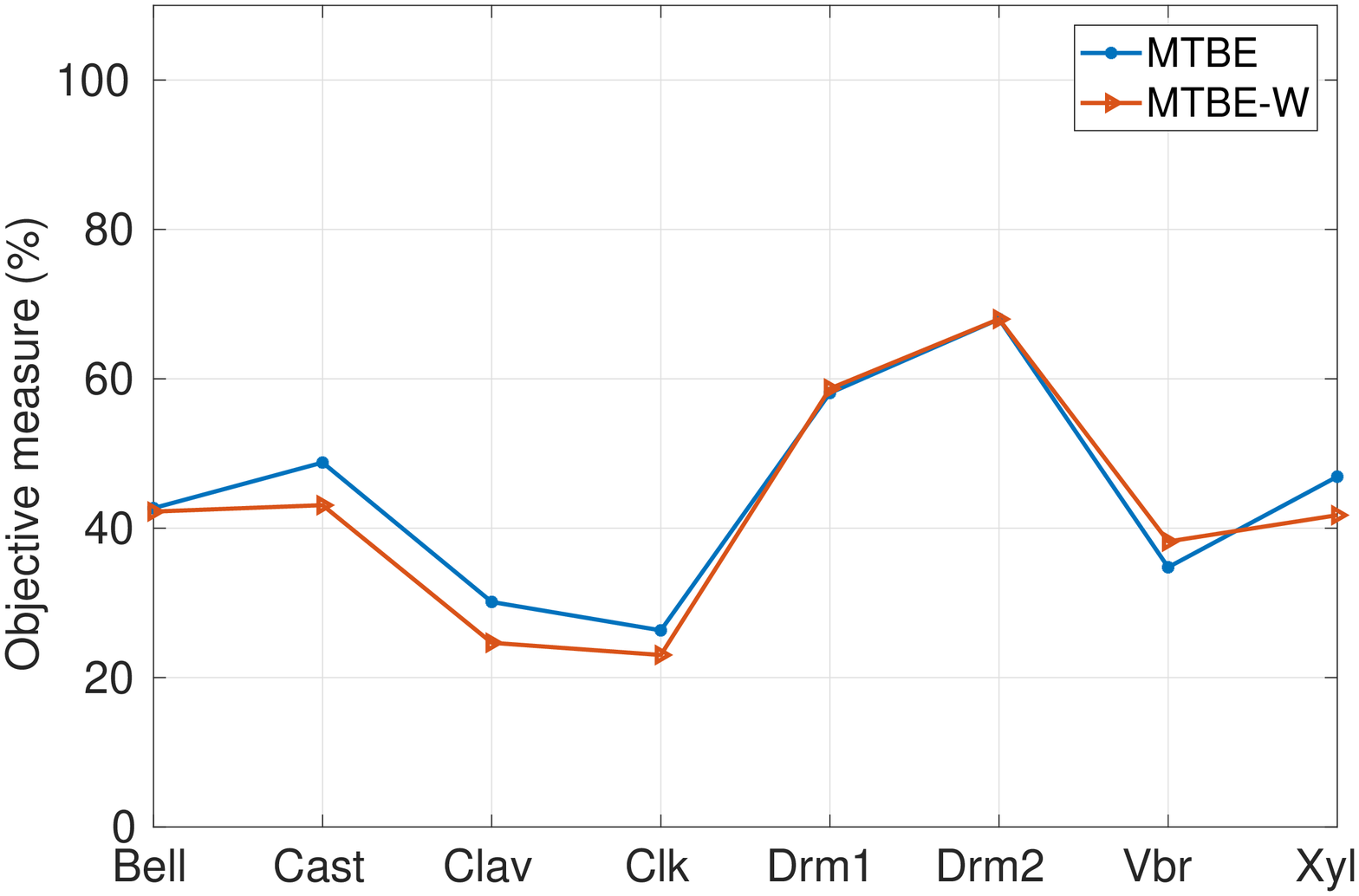}
 \end{minipage}
 \caption{{ (Color Online) Perceptual score and objective  scores for ensemble source width (ESW)}}
 \label{fig_perceptionA}
 \end{figure*}

In each experimental trial, the listener is presented with  musical instrument stimuli of one row as shown in Table.\ref{table_musical_instruments}. In each experimental trial,  the listener is asked to rate the perceived source width expansion in comparison to the explicit reference stimuli of minimum R10 and maximum R100. High pass stimuli with cut-off frequency at 8000 Hz, (8k-20k Hz) is taken as R10, played through the third single loudspeaker (almost center for the listener). Wide band noise, low pass filtered up to 20 kHz, played through 6 loudspeakers (of a physical width of 150cm) is chosen as R100. The GUI for the experiment is shown in Fig.\ref{fig_gui}. Here RMin=10\% and RMax=100\% are explicitly shown. We also provide the narrow stimuli (single channel of the instrument) to the listeners to perceive the difference between the widened stimuli and the original narrow stimuli for each instrument (NS in the GUI). The test stimuli are randomized on the panel and the listener is asked to rate the perceived width of each test stimuli using the scale of range  0-120\%. The listener has the choice to rate even greater than the reference R100 and below the reference R10 if s/he perceives the width wider or narrower, respectively, with respect to the given references. The perceptual rating, given by three listeners is shown in Fig.\ref{fig_perceptionA}. We observe that there is difference among listeners. Listeners L1 and L2 use a wider perceptual scale than L3.

\vspace*{-3mm}
\section{Mean Time-bandwidth Energy (MTBE)} \label{sec_mtbe}
Motivated by spatio-gram that can be effective for sounds of time-varying spectral nature, we study the nature of spectro-temporal patterns aiding source width expansion. We perform short-time Gabor filter bank analysis of music instrument sounds. In order to determine how closely the filter banks need to be placed, we study the frequency discrimination thresholds\cite{wier1977frequency} and observe that the frequency discrimination thresholds are nearly linear as a function of frequency \cite{moore2012introduction}\cite{wier1977frequency}. We quantize the frequency range into three bins of low, mid and high frequencies and use a resolution $\sim$10 times the maximum thresholds at low and mid frequencies and  $\sim$5 times in the case of high frequencies for ease of  computation. The filters are densely placed  in the low frequency region upto 800 Hz separated by 10 Hz, upto 5 kHz separated by 100 Hz and upto 16 kHz separated by 500 Hz. We use Gabor filters with a time-spread ($T_k$) of inverse of the critical bandwidth (CB) at the center frequency ($\omega_k$); the center frequency is mapped to a linear interpolation of the experimentally determined CBs in Zwicker's work \cite{zwicker1980analytical}.
\par 
Let $s[n]$ be  a monophonic music signal. We compute the  energy in a spectro-temporal patch and determine the mean time-bandwidth energy (MTBE) for each instrument as follows. Here, let $w_k[n]$ be cosine modulated Gaussian windows for Gabor filtering.
\begin{align}
 w_k[n] &= e^{-\frac{(t-t^0_k)^2}{2T^2_k}} \cdot cos{\omega_kn} \\
 y[k,n] &= s[n].w_k[n] 
 \end{align}
 The summation time-period ($T_k$) varies inversely depending on the corresponding frequency resolution of the filter $w_k[n]$. 
 \begin{align}
 E[k,m_k] &= \sum_{n=1}^{T_k} \big\| y[k,(m_k-1)T_k+n] \big\|^2 
 \end{align}
 The mean time-bandwidth energy is estimated as follows:
 \begin{align}
 E_M  &= \frac{1}{K}  \sum_{k=1}^{K} \frac{1}{M_k} \sum_{m_k=1}^{M_k} 10\log_{10} (E[k,m_k])
\end{align}
Here, K is the total number of filters over the full bandwidth. We also compute the mean weighted time-BW energy (MTBE-W) with higher weighting for low and high frequencies and low weights for mid range of frequencies.  The relative weights are chosen based on IACC based experiments reported earlier \cite{mason2005frequency}.  
\par 
We obtain the relative MTBE scores (in \%) by normalizing the MTBE values with maximum perceptual score of listener L1 for each instrument class, as listener L1 uses the wider range of perceptual scale. 
\par 
We observe that low pitch sustained instruments  create wider percept and transients have the lowest average score across instruments. In string instruments, low pitched Grand piano, Guitar create wide perception on an average. Low pitch male voice creates wide percept in L1 and L2. In percussion instruments, sustained drum creates wider percept than transient drum in all listeners. In L2, sustained bell and Glokenspeil give rise to width perception. Among wind instruments, low pitched and sustained tuba and saxaphone  create wide percept. 
\par
In all instruments, the objective measure with logirthmic weights spans a compressed dynamic range than the perceptual score. In case of string instruments, the relative rating of Cello, Grand piano, Guitar, Harp are predicted for most of  L1 and L2 rating. In case of viola and violins, the relative order does not match. This could be because of high variation in the style of rendering of violin-1 and violin-2 that listeners could have focused on cues at different instants and rated the ESW \cite{arthi2019multi}. Perceptually Guitar and grand piano create widest percept in L1 and L2 respectively but objective measure predicts organ. Harp is rated perceptually wider than piano but objective measure contradicts.  Fine spectro-temporal analysis may be required to predict these perceptual scores. 
\par 
In case of vocals, objectively low pitched  male rendering are wider than female vocals. L1 rates the bass male as the widest.  High pitched soprano 2 rendering has interfering low pitch accompany instrument which contributes to perceptual score but listeners follow the high pitched vocal that this objective measure can be omitted for analysis. Relative order of Tenor-M with respect to Alto-female and soprano-1 female for L2 are predicted by the objective measure .  Male speech is perceived wider than female speech in L1 and L2 and it is predicted by the objective measure. Vocals are perceived wider than speech  by all listeners but the objective measure does not reflect that. The objective measure appears to have limitation when comparing high varing styles of rendering, as observed in case of violins . 
\par
In case of wind instruments, most of the relative perceptual order of clarinet, flute, horn and oboe for L1 are predicted in the objective measure. Similarly the perceptual relative orders of sax, trombone, trumpet and tuba are predicted by the objective measure. Low pitch weighted objective measure is closed to perceptual measure of tuba for L1. Thus, perceptual weighting for objective measure is required for low-pitched instruments.  High pitched piccolo is estimated to have lowest in the objective measure but the perceptual loudness associated with high pitched sounds may ve influenced the perceptual ESW, though the stimuli are normalized by loudness in SPL pressure levels. This could be a reason for discrepancy  in rating for piccolo. 
\par
In case of percussion instrument, most of the relative perceptual order of bell, glockenspiel, transient drum (drm-1), sustained drum (drm-2), vibraphone and xylophone match for L1. In case of L2, relative rating of drm-1, drm-2, vibraphone and xylophone match. High perceptual rating for bell and glockenspiel suggest L2 gives more weight for the sustained portion of the transients. Claves is always rated wider than castanet but objective measure proposes otherwise. More analysis is required in this direction. 
\par
In summary, this objective measure correlated well with sustained rendering of wind instruments and transient instruments like sustained drum; and does not correlate well with stringed transient instruments like organ and high varying styles of violin. Also, fine differences between  vocals and speech is not predicted by this objective measure. Fine time-varying analysis is required to  bring in  understanding the perceptual weight of fine AM-FM variations in spatialisation. 

\section{Summary Discussion}  \label{sec_discussion}
We discuss here the results of POSC, short-term POSC or spatiogram, MTBE and the perception experiment results
\begin{enumerate}[wide = 0pt]
 \item {We observe that each source is well localized with POSC function domain compared to magnitude based correlation function (MSC). This shows that the binaural phase-difference plays a key role in the spatialisation or spatial perception of multi-source perception and is of paramount importance. While algorithms in hearing applications \cite{doclo2005extension}\cite{marquardt2013coherence}\cite{itturriet2019perceptually} and Directional coding \cite{hotho2008multichannel}\cite{breebaart2005mpeg} are optimized to retain ITD, ILD and IC, it may be more important to analyse and  preserve the POSC function, which is a combination of ITD, ILD and IC. In-fact, the spatiogram combines all the three parameters and captures the multiple sources present in a time-varying sense. Hence, while rendering, we can develop algorithms to optimize to preserve spatio-gram signature for multi-source ensemble rendering. }
\item {For spatial continuum applications, many techniques of HRTF interpolation have been developed where interpolation is done in  time domain or frequency domain \cite{sreenivas2000head}. We can observe that POSC offers a domain for interpolation of HRIRs, where we can optimize the interpolation technique to retain the spatial continuum in the POSC  domain, retaining the continuum of binaural phase difference signature. }
\vspace*{-2mm}
\item{Even in the case of reverberation in concert-hall and room acoustics studies, it is observed that IACC based objective measures fail to capture perceptual source widening completely \cite{morimoto2005appropriate}. 
An objective measure for ASW and LEV of an arbitrary signal is still an open question \cite{morimoto1995practical} \cite{morimoto2005appropriate} \cite{bomer2011effect}.  We can improve the present POSC estimation to explore the direction of arrival (DOA) estimate for attenuated components of early and late reverberation and develop a similar method of decomposing the objective measure into 3 parts (i) estimate timbre-independent angular distribution (ii) weight for reverberation strength (iii) timbre-dependent perceptual weight. In this approach the estimation of angular spread is a challenge because of attenuated early and late components, masked by the direct component}
\vspace*{-2mm}
 \item {The HRTF correlation filters introduce spurious peaks and hence  may get detected as presence of actual sources. Also, in this work, we have assumed the number of sources as a known parameter. In actual distributed source and multi-loudspeaker rendering scheme, we may not know the actual number of sources. Hence, automatic detection of the presence of actual sources and edge sharpening need algorithmic improvement.} 
 \item {ILD plays a significant role in the relative peak strengths of multiple sources. Direction dependent ILD compensation may be required in-order to determine the relative strengths of the individual sources}
 \item {In case of ensemble presentation, the edges determine the spatial extent. Hence, edge detection in the POSC function is important. With head rotation, the peak levels of the edges change depending on the direction of location of the sources and the relative head movement. This could be a significant cue for edge detection in presence of noise or interference or  reverberation. Time-varying POSC estimation with head movement could improve the angular estimation and orientation. }
 \vspace*{-2mm}
 \item {Most of the computational models for localization and/ or width perception are based on time-delay based Jeffress model of correlation \cite{jeffress1948place}\cite{hirvonen2006perception}. We observe from the POSC function that an additional HRIR phase matching block is required after the phase only correlation for multiple source localization. This opens up new domains of binaural computational models for multiple/mixed source localization.}
 \vspace*{-2mm}
 \item {From the objective measure, we observe that MTBE predicts the perceptual score in sustained cases in Fig.\ref{fig_perceptionA}. Also, there are significant differences between MTBE and perceptual measure. This missing difference could be because of fine spectro-temporal AM-FM variations and it may be required to model and include fine AM-FM variations in the computational model along with weighting function.}
\end{enumerate}
\vspace*{-2mm}


\section{Conclusion} 
We have analytically and experimentally developed an objective measure for the perception of ensemble source width (ESW) using a combination of two factors (i) Timbre independent directional angular measure (ii) Timbre depended perceptual weight for the overall spatial measure, called mean time-bandwidth energy (MTBE). We can extend this objective measure for ESW to auditory source width perception in concert hall acoustics also. We have also developed a time-varying spatial correlation function called spatio-gram to analyse relative motion of sources with respect to listener and head movement.  

\vspace*{-3mm}
\bibliography{references_v1.bib}

\end{document}